\documentclass[twoside,twocolumn]{article}

\usepackage{mathrsfs} 
\usepackage{blindtext} 
\RequirePackage{amsgen}
\usepackage[sc]{mathpazo} 
\usepackage[T1]{fontenc} 
\usepackage{microtype} 
\usepackage{amsmath}
\usepackage{amssymb}
\usepackage[english]{babel} 
\usepackage{xcolor}
\usepackage[hmarginratio=1:1,top=32mm,columnsep=20pt]{geometry} 
\usepackage[hang, small,labelfont=bf,up,textfont=it,up]{caption} 
\usepackage{booktabs} 
\usepackage{physics}

\usepackage{lettrine} 
\usepackage{tabularx}
\usepackage{enumitem} 
\setlist[itemize]{noitemsep} 

\usepackage{abstract} 

\usepackage{titlesec} 
\renewcommand\thesection{\Roman{section}} 
\renewcommand\thesubsection{\roman{subsection}} 
\titleformat{\section}[block]{\large\scshape\centering}{\thesection.}{1em}{} 
\titleformat{\subsection}[block]{\large}{\thesubsection.}{1em}{} 

\usepackage{fancyhdr} 
\usepackage{titling} 

\usepackage{hyperref} 

\pretitle{\begin{center}\Huge\bfseries} 
\posttitle{\end{center}} 
\title{Reduction of Lagrangian Equations of Motion of Modified Newtonian Theory of Gravity with respect to the Similarity Group} 
\author{%
\textsc{Sahand Tokasi}\thanks{sahand.tokasi@uni-tuebingen.de} \\[1ex] 
\normalsize University of T\"ubingen \\ }
\date{\today} 
\renewcommand{\maketitlehookd}{%
\begin{abstract}
\noindent 
The equivalence class of absolute configurations of a system under the group of similarity transformations $Sim(3)$ is called the shape of the system. The $Sim(3)$ invariant Lagrangian of the modified Newtonian theory ensures the existence of its law of motion on shape space. To deduce the equations of motion for a system's shape degrees of freedom from its evolution equations for the $3N$ absolute configuration degrees of freedom, the Boltzman-Hamel equations of motion in an non-holonomic frame on the tangent space $T(Q)$ to the system's absolute configuration space $Q$ is adapted to the $Sim(3)$ fiber bundle structure of the configuration space. The derived equations of motion on shape space enable us, among other things, to predict the evolution of the shape of a classical system governed by this theory without any reference to its absolute position, orientation, or size in space. The paper will explain, that by treating the measuring instruments as part of the matter in the theory, how the mass metric $\textbf{M}$ on the configuration space $Q$ uniquely defines a metric on the reduced tangent bundle $\frac{T(Q)}{Sim(3)}$, and how the unique metric structure on shape space $S$ can be derived.
After deriving the reduced equations of motion on shape space for a general $N$-body system, the shape equations of motion for a three-body system in suitable coordinates is given as an illustration. \newline

\centering "Dedicated to the good memories of my father, Mohammadreza Tokasi."\newline

\end{abstract}
}

\begin{document}

\maketitle 

\tableofcontents
\section{Introduction}
In \cite{Me4} and \cite{8}, a classical (non-relativistic) theory of gravity based on relational ideas in physics has been developed. This so-called Modified Newtonian Theory (or MNT for short) has a similarity invariant potential function $V:Q\rightarrow \mathbb{R}$ and hence the similarity group $Sim(3)$ constitutes its symmetry group. The MNT equations of motion given in Section (II.i)\footnote{For any of the possible functions $G$ discussed there.} of \cite{Me4} are mathematically like the equations of motion of (original) Newtonian theory, a set of coupled second order partial differential equations that mixes the evolution of the $3N-7$ relational (or shape) degrees of freedom of the system with its $7$ absolute (or gauge or collective) degrees of freedom. In this paper, we aim to decouple the evolution of the two aforementioned sets of degrees of freedom and, in this way, derive the equations governing the evolution of the shape degrees of freedom. In other words, given the Lagrangian equations of motion of the modified Newtonian theory on absolute configuration space $Q\cong \mathbb{R}^{3N}$, we seek the equations of motion of MNT on shape space $S=\frac{Q}{Sim(3)}$. \newline\newline
Because MNT has the entire similarity group $Sim(3)$ as its symmetry group, the extensive literature on the reduction of equations of motion with respect to symmetry groups (\cite{1},\cite{2},\cite{3},\cite{7},\cite{LPE},\cite{MarsdenRatiuScheurle},\cite{MestdagCrampin},...) assists us to find the answer. As the constancy of angular momentum forms an anholonomic constraint on $T(Q)$, the formulation of Lagrangian systems on non-holonomic frames and their reduction with respect to symmetry groups(\cite{1},\cite{2},\cite{3},\cite{7}) are used and extended (to include the scale transformations) to find the reduced equations of motion of MNT on shape space $S$. The reduced equations of motion w.r.t. a symmetry group are also known as Lagrange-Poincare equations (\cite{LPE},\cite{MarsdenRatiuScheurle},\cite{MestdagCrampin}).\newline\newline
In Section (II), we extend the geometric setting on the center of mass configuration space $Q_{cm}\cong\mathbb{R}^{3N-3}$ of an $N$-particle system considered as a $SO(3)$ fiber bundle, as explained in \cite{1}, \cite{2} and \cite{3}, to scale transformations $Sc$ and develop the construction of the $Sim(3)$ fiber bundle. Under the action of the group of scale transformations $Sc$, the internal configuration space $Q_{int}:=\frac{Q}{E(3)}$ becomes a fiber bundle whose base space is called shape space $S:=\frac{Q_{int}}{Sc}=\frac{Q}{Sim(3)}$. We explain in Section (II.ii) how the mass metric $\textbf{M}$ on the absolute configuration space $Q\cong \mathbb{R}^{3N}$ induces a unique metric $\textbf{N}$ on the reduced tangent bundle $\frac{T(Q)}{Sim(3)}$ using the principle of relationalism. We also present here a way to solve the non-uniqueness problem in the metric $N$ on shape space, which aroused in the DGZ-approach\cite{DSN}, by explaining the physical origin of the conformal factor. In particular, we explain that once the behavior of the measuring units made out of matter is theoretically included, the \textit{unique metric structure on Shape space} reveals itself and clarifies the relationship between the choice of a unit of length and the choice of a conformal factor and elaborate that all reasonable choices of length units lead to the same metric on shape space. In Section (II.iii), we develop representations of the group of similarity transformations $Sim(3)$ and its Lie-algebra $\textbf{sim}(3)$. Considering $Q_{int}$ as a $Sc$ fiber bundle, we define a new connection form on $Q_{int}$ and show by an explicit calculation that it is a flat connection form.\newline
In Section (III), after reviewing the Lagrangian formulation of mechanics in non-holonomic frames and its Boltzmann-Hamel equations of motion, we will derive the equation of motion for the shape degrees of freedom of an $N$-particle system whose behavior in absolute space and time is given by the \textit{modified Newtonian theory}. As explained in \cite{Me4}, the direct implementation of the principle of relationalism leads to scale invariant interaction potentials, which is the key to the complete decoupling of the dynamics on shape space from the gauge degrees of freedom, i.e., the $Sim(3)$ degrees of freedom.
\newline
In the end, as an illustration of the extended formalism, in Section (IV), we explicitly derive the shape equations of motion of a modified Newtonian three-body system in suitable coordinates.
\section{Ingredients for the Reduction with respect to the similarity group}
Which equations of motion do the evolution of the shape degrees of freedom of a modified Newtonian system satisfy is the central question to be answered in this paper. To this end, we will explain in this section how to derive the necessary ingredients, i.e., the metric $\textbf{N}$ on the reduced tangent bundle $\frac{T(Q)}{\mathcal{A}_{Sim(3)}}$ and the connection form $\omega$ for the $Sim(3)$ fiber bundle.    
The mass metric \textbf{M} of the absolute configuration space $Q\cong \mathbb{R}^{3N}$ induces metrics on the reduced spaces, such as the internal configuration space $Q_{int}=\frac{Q}{E(3)}$, shape space $S=\frac{Q}{Sim(3)}$, and the $Sim(3)$-reduced tangent bundle in a natural fashion. In the subsection (II.i) following \cite{1},\cite{2}, we first review the derivation of the metric $B$ on $Q_{int}=\frac{Q_{cm}}{SO(3)}$. Then, using the principle of relationalism, we explain a new way to derive a metric $\textbf{N}$ on the $Sim(3)$-reduced tangent bundle $\frac{T(Q)}{Sim(3)}$ from the mass metric $\textbf{M}$ on the absolute configuration space. At last, we present our derivation of the unique metric structure $N$ on shape space $S$.  
\subsection{Metric on the internal space $Q_{int}$}
The absolute configuration space of a $N$-particle system is the set \[Q=\{x=(\pmb{x}_{1},...,\pmb{x}_{N})\mid \pmb{x}_{i}\in\mathbb{R}^{3}\}\cong \mathbb{R}^{3N}\] 
The center of mass system is the following subset of $Q$ \[Q_{cm}=\{x=(\pmb{x}_{1},...,\pmb{x}_{N})\mid \sum^{N}_{j=1}m_{j}\pmb{x}_{j}=0\}\cong \mathbb{R}^{3(N-1)}\] with $\pmb{x}_{j}\in \mathbb{R}^{3}$. A new coordinate system on $Q$ adapted to the projection $Q\rightarrow Q_{cm}$ is given by the following linear transformation
\[\left(\begin{array}{c}\pmb{x}_{1}\\ \pmb{x}_{2}\\ .\\.\\.\\ \pmb{x}_{N}\end{array}\right)\rightarrow \left(\begin{array}{c}\pmb{r}_{1}\\ \pmb{r}_{2}\\ .\\.\\.\\\pmb{r}_{N-1}\\ \textbf{R}_{cm}\end{array}\right)\]
where $\pmb{R}_{cm}=\frac{1}{\sum_{i=1}^{n}m_{i}}\sum_{\alpha=1}^{n}m_{\alpha}\pmb{r}_{\alpha}$ is the center of mass of the system, and the $N-1$ vectors $\pmb{r}_{i}$ are the mass-weighted Jacobi vectors defined as follows 
\begin{equation}\label{Jacobi}
\pmb{r}_{j}:=(\frac{1}{\mu_{j}}+\frac{1}{m_{j+1}})^{-\frac{1}{2}}(\pmb{x}_{j+1}-\frac{1}{\mu_{j}}\sum^{j}_{i=1}m_{i}\pmb{x}_{i})
\end{equation}
with $\mu_{j}:=\sum^{j}_{i=1}m_{i}$. Then, the center of mass configuration space $Q_{cm}$ can be expressed in these coordinates by 
\begin{equation}\label{jacocm}
Q_{cm}\cong \{x=(\pmb{r}_{1},...,\pmb{r}_{N-1})\mid \pmb{r}_{j}\in \mathbb{R}^{3}, j=1,...,N-1\}
\end{equation}
By the introduction of the previous coordinate transformation $(\pmb{x}_{1},...,\pmb{x}_{N})\rightarrow (\pmb{r}_{1},...,\pmb{r}_{N-1},\textbf{R}_{cm})$ on the absolute configuration space $Q$, the system's total kinetic energy naturally separates in the center of mass kinetic energy and the center of mass rotational energy naturally. Specifically system's total kinetic energy 
\[K=\frac{1}{2}\sum_{\alpha=1}^{N}m_{\alpha}\mid\dot{\pmb{x}}_{\alpha}\mid^{2}=\frac{1}{2}\sum_{\alpha,\beta=1}^{N}\textbf{M}_{\alpha\beta}(\dot{\pmb{x}}_{s\alpha}.\dot{\pmb{x}}_{s\beta})\] with $\textbf{M}_{\alpha\beta}$
being the $3N\times 3N$ kinetic tensor (\ref{KEN}), transforms to
\begin{equation}\label{transredmet}
K=\frac{1}{2}\sum_{\alpha=1}^{N-1}\mid\dot{\pmb{r}}_{\alpha}\mid^{2}+\frac{\sum_{i=1}^{n}m_{i}}{2}\mid\dot{\mathbf{R}}_{cm}\mid^{2}
\end{equation}
The center of mass configuration space $Q_{cm}$ forms a stratified fiber bundle by the action of the group $G=SO(3)$ on it.
Let us review how the metric $B$ on the internal space $Q_{int}=\frac{Q_{cm}}{SO(3)}$ can be derived from the $SO(3)$ invariant mass metric $\textbf{M}$ on the the center of mass configuration space $Q_{cm}$, i.e.,
\begin{subequations}
\begin{align}
       \textbf{M}_{x}(u,v)=\sum m_{k}<\pmb{u}_{k}\mid \pmb{v}_{k}>\\
       \textbf{M}_{x}(u,v)=\textbf{M}_{gx}(gu,gv)
\end{align}
\end{subequations}
where $u=(\pmb{u}_{1},...,\pmb{u}_{N})$ and $v=(\pmb{v}_{1},...,\pmb{v}_{N})$ are members of $T_{x}(Q_{cm})$, so being any two tangent vectors of $Q_{cm}$ at the point $x\in Q_{cm}$. \newline
Given two internal vectors \[v',u'\in T_{q}(Q_{int})\] there are unique vectors $u,v\in T_{x}(Q_{cm})$ \footnote{Namely their horizontal lifts} so that 
\[\left\{
  \begin{array}{lr}
    \pi(x)=q\\
    \pi_{*}(u)=u'\\
    \pi_{*}(v)=v'
  \end{array}
\right.
\] 
Now, the metric $B$ on $Q_{int}$ can be defined by the following equation:
\begin{equation}\label{B metric}
B_{q}(v',u'):=\textbf{M}_{x}(v,u).
\end{equation}
As the metric $\textbf{M}$ is $SO(3)$ invariant, to which $x\in \pi^{-1}(q)$ the internal vectors $v,u$ are lifted, would not make any difference for the value assigned by $B_{q}$. Hence, the metric $B$ is well-defined.

\subsection{Unique metric on $Sim(3)$-reduced tangent bundle}
Considering that the measurement of velocities is essentially an experimental task, the transformation law of velocities under scale transformations of the system (or any other system transformation) must also involve experimental considerations. Based on the principle of relationalism, we have shown that the behavior of rods and clocks under scale transformations of the system is such that the measured velocities of objects (subsystems) are invariant under scale transformations. This invariance is a natural consequence of the simultaneous expansion of the measuring rod and the corresponding dilation of the unit of time (see Section (II) of \cite{Me4} for an explanation of this fact). Hence, a velocity vector \[v_{x}=(v_{1},...,v_{N})\in T_{x}(Q_{cm})\] of an N-particle system transforms under scale transformations of the system
\[x\rightarrow b x\]
as follows
\[v_{x}=(v_{1},...,v_{N})\in T_{x}(Q_{cm})\] \[\downarrow \] \[ v_{b x}=(v_{1},...,v_{N})\in T_{c x}(Q_{cm})\]  
Given the above action $\mathcal{A}_{b}$ of $b\in Sc\subset Sim(3)$ on velocities(or on $T(Q)$); the mass metric is a $\mathcal{A}_{Sc}$ invariant metric on $Q$, as can be seen by a short calculation:
\begin{equation}\label{siip}
    \textbf{M}_{x}(v_{x},u_{x})\rightarrow  \textbf{M}_{b x}(\mathcal{A}_{b}v_{x},\mathcal{A}_{b}u_{x})= \textbf{M}_{x}(v_{x},u_{x}) 
\end{equation}
where the equalities $\textbf{M}_{x}=\textbf{M}_{bx}=\textbf{M}$ and $\mathcal{A}_{b}v_{x}=v_{x}$ has been used. Considering $T(Q_{int})=T(\frac{Q}{E(3)})$ as a $\mathcal{A}_{Sc}$ fiber bundle, the mass metric $B$ on $Q_{int}=\frac{Q}{E(3)}$ (defined previously by expression (\ref{B metric})) induces a unique metric \[\textbf{N}_{s}:T(Q_{int})/\mathcal{A}_{Sc}\times T(Q_{int})/\mathcal{A}_{Sc}\rightarrow \mathbb{R}\] 
as follows
\begin{equation}\label{shapemetric1}
\textbf{N}_{s}(v',u'):=B_{q}(v,u)
\end{equation}         
where
\[\left\{
  \begin{array}{lr}
    \pi(q)=s\\
    \pi'(u)=u'\\
    \pi'(v)=v'
  \end{array}
\right.
\] 
with the projection maps defined as follows \[\pi:Q_{int}\rightarrow S\]
\[\pi':T(Q_{int})\rightarrow T(Q_{int})/\mathcal{A}_{Sc} \]
Since the above construction is $\mathcal{A}_{Sc}$ invariant, to which $q\in \pi^{-1}(s)$ the pair of shape vectors $v',u'\subset T_{q}(Q_{int})/\mathcal{A}_{Sc}$ is lifted, does not make any difference for the value assigned by $\textbf{N}_{s}$ to them. Therefore, the metric \textbf{N} is well defined. Moreover, this method brings one uniquely to the shape (kinetic)metric $\frac{1}{2}\textbf{N}$ on $T(Q_{int})/\mathcal{A}_{Sc}$ without the need to introduce a conformal factor and the ambiguity involved with it.
\subsection{Unique metric on shape space}
Denote the push forward of vectors under the group of scale transformations by $Sc_{*}$, i.e.,
\[Sc_{*}:T_{q}(Q)\rightarrow T_{bq}(Q)\]
for any $b\in\mathbb{R}^{+}$ representing a member of the group $Sc$. 
Since the mass metric $\textbf{M}$ is not scale invariant, i.e.,
\begin{equation}\label{nsi}
    \textbf{M}_{b x}(Sc_{*}u,Sc_{*} v)=\textbf{M}_{b x}(b u,b v)=b^{2} \textbf{M}_{b x}(v,u)
\end{equation}
\[=b^{2} \textbf{M}_{ x}(v,u)\neq \textbf{M}_{ x}(v,u) \]
it is generally believed that, unlike on $Q_{int}$, it does not uniquely induce a metric on shape space $S$. It is explained in \cite{DSN} that one can introduce a new $Sim(3)$ invariant metric on $Q$, which then induces a metric on shape space in a natural way. As the mass metric $\textbf{M}$ is already rotation- and translation invariant, the easiest way to arrive at a similarity invariant metric is to multiply the mass metric by a function $f(x)$ (the so-called conformal factor) so that the whole expression \[\textbf{M}'_{x}:=f(x)\textbf{M}_{x}\] becomes scale invariant, i.e., 
\[\forall b\in \mathbb{R}^{+}, \forall u,v \in T_{x}(Q) :\] \[\textbf{M}'_{b x}(Sc_{*} u,Sc_{*} v)=f(bx)\textbf{M}_{b x}(Sc_{*} u,Sc_{*} v)\]\[= f(x)\textbf{M}_{x}(u,v)=\textbf{M}'_{x}(u,v)\]
Note that the function $f$ must be translation and rotation invariant so that it does not spoil the Euclidean invariance of the mass metric. As $\textbf{M}'_{x}=f(x)\textbf{M}_{x}$ is now a metric invariant under the whole similarity group, it induces a metric $N$ on shape space as follows
\begin{equation}\label{shapemetric}
N_{s}(v',u'):=\textbf{M}'_{x}(v,u)=f(x)\textbf{M}_{x}(v,u)
\end{equation}         
where
\[\left\{
  \begin{array}{lr}
    \pi(x)=s\\
    \pi_{*}(u)=u'\\
    \pi_{*}(v)=v'
  \end{array}
\right.
\] 
with the projection map $\pi:Q_{cm}\rightarrow S=\frac{Q}{sim(3)}$.\newline
When the action of scale transformation on $T(Q)$ is defined by the differential of the scale transformations, i.e., $Sc_{*}$, from the behavior of the mass metric $\textbf{M}$ under scale transformation (\ref{nsi}) one sees that any rotation- and translation invariant homogeneous function\footnote{ A function of $r$ variables $x_{1},...,x_{r}$ is being called homogeneous of degree $n$ if $f(c x_{1},...,c x_{r})=c^{n}f(x_{1},...,x_{r})$,$\forall c$ } of degree $-2$  perfectly meets all the requirements of a conformal factor. For instance
\begin{equation}\label{cf1}
    f(x)=\sum_{i<j}\mid\mid \pmb{x}_{i}-\pmb{x}_{j}\mid\mid^{-2}
\end{equation}
or 
\begin{equation}\label{cf2}
    f(x)=I^{-1}_{cm}
\end{equation}
 where 
  \[I_{cm}(x)=\sum_{j}m_{j}\mid\mid \pmb{x}_{j}-\pmb{x}_{cm}\mid\mid^{2}\]\[=\frac{1}{\sum_{i}m_{i}}\sum_{i<j}m_{i}m_{j}\mid\mid \pmb{x}_{j}-\pmb{x}_{i}\mid\mid^{2}\] 
 are two legitimate examples of conformal factors (\cite{DSN}). \newline\newline
 Since introducing different conformal factors leads to different metrics on shape space, the above derivation of the shape metric suffers from a non-uniqueness problem. This arbitrariness in the metric of shape space, caused by the arbitrariness in the choice of a conformal factor, is, in our opinion, due to the forgotten connection of length measures with the real rulers. We propose another way of deriving the metric on shape space below, in which the mentioned problem does not arise. In particular, 
 we explain why the measured mass metric is on its own scale invariant. A conformal factor would be required if we had access to absolute rulers and could thus measure absolute lengths. Since all rulers are themselves subsystems of the universe, they are also subject to the transformations applied to the universe. Taking this physical fact into account resolves the mentioned problem and provides us with the unique metric $N$ on shape space. In other words, when one uses units of measurement formed from matter instead of absolute units of measurement, one finds that the measured mass metric induces a unique metric on shape space. This way, we also explain the relationship between length units and conformal factors.\newline\newline
As mentioned earlier, Mathematically, a metric $\textbf{G}$ on a manifold $Q$ is called scale invariant if and only if
\begin{equation}\label{mms}
    \forall v_{1},v_{2}\in T_{q}(Q): \textbf{G}_{q}(v_{1},v_{2})=\textbf{G}_{bq}(Sc_{*}v_{1},Sc_{*}v_{2})
\end{equation}
where $Sc_{*}:T(Q)\rightarrow T(Q)$ denotes the push forward of vectors along the scale transformations $Sc:q\rightarrow bq$ on $Q$. Since $Sc_{*}v=bv$, we saw that the mass metric $\textbf{M}$ is not scale invariant in this sense(\ref{nsi}). However, what one physically measures and is relevant is not $\textbf{M}$ but
\begin{equation}\label{pmm}
    \textbf{M}^{(m)}_{q}(v_{1},v_{2})=\frac{\textbf{M}_{q}(v_{1},v_{2})}{\textbf{M}_{q}(\pmb{q}_{i}-\pmb{q}_{j},\pmb{q}_{i}-\pmb{q}_{j})}
\end{equation}
where $1<i,j<N$ are two particles used to define the length unit. It is another way to see how the previously criticized arbitrariness of the metric on shape space disappears by using real units of measurement instead of ``inaccessible absolute units''. The \textbf{measured mass metric} is on its own scale invariant in the mathematical sense mentioned above. One could say that part of the arbitrariness of the conformal factor is now actually shifted to the arbitrariness in the choice of a length unit, i.e., which particles $i$ and $j$ one chooses to define the unit of length. However, one should realize that all reasonable choices for the length unit will result in the same metric $N$ on shape space. A reasonable choice for the length unit would not lead to any fictitious forces. For instance, choosing two particles forming a harmonic oscillator (w.r.t. the gross background structure of the universe) as the length unit would be a terrible choice. This fact also explains the appearance of non-physical forces encountered by various conformal factors in the DGZ-approach of \cite{DSN}.     \newline
In this regard, the mathematical definition of the notion of scale-invariance in Riemannian geometry (\ref{mms}) is less relevant from the physical point of view. It is due to the use of the differential of the scale transformation $Sc$ as the action of $Sc$ on $T(Q)$ and its decoupling from the physical theory. Here we define a new notion that is more relevant to physics. A metric $G$ on the configuration space $Q$ is called \textit{mechanical similarity invariant} if and only if
\begin{equation} \forall v_{1},v_{2}\in T_{q}(Q), G_{q}(v_{1},v_{2})=G_{bq}(b^{\frac{k}{2}}v_{1},b^{\frac{k}{2}}v_{2})
\end{equation}
where $k$ is the degree of homogeneity of the potential function of the physical theory. The factor $b^{\frac{k}{2}}$ results from the combined effect of the time transformation required by the theory's mechanical similarity \footnote{Mechanical similarity says that if $x(t)=\big(\pmb{x}_{1}(t),...,\pmb{x}_{N}(t)\big)$ is a solution to the $N$-body problem with a homogeneous potential function $V$ of degree $k$, then $x'(t')=\big(b\pmb{x}_{1}(t'),...,b\pmb{x}_{N}(t')\big)$ is also a solution of the theory with $t'=tb^{1-\frac{k}{2}}$. This type of time transformation was required for the dynamical equivalence of the two alternative universes as explained in \cite{Me4}.}, and the ruler's extension. Thus, instead of defining the action of $Sc$ on $T(Q)$ by push forward $Sc_{*}$, we define the action of $Sc$ by the mechanical similarity transformation on $T(Q)$. The mass metric $\textbf{M}$ is mechanical similarity invariant for the modified Newtonian theory but not for the original Newtonian theory. A mechanical similarity invariant metric defines a unique metric $\textbf{N}$ on the $Sim(3)$-reduced tangent bundle $\frac{T(Q)}{Sim(3)}$.\newline\newline
At last, we note that $N$ is a metric on $T(S)=T(Q_{int})/Sc_{*}$, while $\textbf{N}$ is a metric on $T(Q)/\mathcal{A}_{Sc}$. Thus, these are metrics on two different vector bundles over $S$. Although we intuitively expect them to represent the same physical entity, their mathematical equivalence is not obvious to us. In the remainder of this text, we will always work with $T(Q)/\mathcal{A}_{Sc}$ and use $\textbf{N}$. With some abuse of notation, we will denote both bundles by $T(S)$, but it is clear from the context which bundle is meant.

\subsection{Connection form for $Sim(3)$ fiber bundle} 
For the derivation of the Lagrangian equations of motion on shape space $S$ (which is the topic of the next section), in addition to a similarity invariant potential function on absolute configuration space (introduced in \cite{Me4},\cite{8}) and the metric $\textbf{N}$ on shape space $S$, we need the appropriate connection form $\omega$ on the absolute configuration space $Q$, compatible with the $Sim(3)$ fiber bundle structure, and an expression of the mass metric in suitable coordinates (or one- forms).\newline
In this section, we discuss some features of the similarity group and present two representations of this group and its Lie-algebra $\textbf{sim}(3)$. We will then use these to construct the connection form of the $Sim(3)$ fiber bundle. At the end of this section, we will show by an explicit calculation that shape space $S$ has the same curvature as the internal configuration space $Q_{int}:=\frac{Q}{E(3)}$.   \newline\newline
\textbf{Construction of the connection form of $Sim(3)$ fiber bundle:}\newline\newline
 Similarity group $Sim(3)$ acts on any point $\textbf{x}\in\mathbb{R}^{3}$ of absolute space as follows
\[\textbf{x}\rightarrow \textbf{x}'=b P\textbf{x}+\textbf{t}\]
where \[b\in \mathbb{R}^{+}\] stands for the spatial scale transformations, \[P\in SO(3)\] for the $3\times 3$ matrix representation of spatial rotations \footnote{which can be parameterized for instance by the three Euler angels.}, and \[\textbf{t}=(t_{1}, t_{2}, t_{3})^{T}\in \mathbb{R}^{3}\] for spatial translations.\newline 
The group of rotations $SO(3)$ does not form a normal subgroup of $Sim(3)$, while the groups of translations $T(3)\cong\mathbb{R}^{3}$ and scale transformations $Sc\cong\mathbb{R}^{+}$ both do. As a result, one recognizes a semi-direct product structure\footnote{Applying a translation and then a rotation is equivalent to applying the rotation and then a translation by the rotated translation vector. Hence $E(3)$ is a semi-direct product of $T(3)$ and $O(3)$, i.e., $E(3)=T(3)\rtimes O(3)$} in $Sim(3)$, i.e.,
\[Sim(3)=Sc\times T(3) \rtimes SO(3)\]
The well-known isomorphism\footnote{see appendix (V.iii) for more details about $R$.}
\[R: \mathbb{R}^{3}\rightarrow \textbf{so(3)}\]
\begin{equation}\label{defR}
R(\pmb{\omega})=\begin{bmatrix}0 & -\omega^{3} & \omega^{2} \\\omega^{3} & 0 & -\omega^{1}\\ -\omega^{2} & \omega^{1} & 0 \end{bmatrix}
\end{equation}
between $\mathbb{R}^{3}$ and the Lie-algebra $\textbf{so}(3)$ of the group of rotations $SO(3)$ will be used by us in the construction of the group representations of $Sim(3)$. \newline
If one thinks of absolute space as a section $\mathbb{R}^{3}\times \{1\}\in \mathbb{R}^{4}$, one can give a representation of the similarity group $Sim(3)$ in terms of the $4\times 4$ matrices of the form 
\begin{equation}\label{SimRep1}
    \begin{bmatrix} b P_{11} & b P_{12} & b P_{13} &  t_{1} \\ b P_{21} & b P_{22} & b P_{23} & t_{2}\\b P_{31} & b P_{32} & b P_{33} & t_{3} \\0 & 0 & 0 & 1 \end{bmatrix}
\end{equation}
The Lie-algebra $\textbf{sim}(3)$ of the similarity group $Sim(3)$ is then given by the matrices
\begin{equation}\label{simRep1}
    \begin{bmatrix} \dot{b} & \omega_{3} & -\omega_{2} & v_{1} \\ -\omega_{3} & \dot{b} & \omega_{1} & v_{2}\\\omega_{2} & -\omega_{1} & \dot{b} & v_{3} \\0 & 0 & 0 & 0 \end{bmatrix}
\end{equation}
where $\pmb{\omega}=(\omega_{1},\omega_{2},\omega_{3}), \textbf{v}=(v_{1},v_{2},v_{3})\in\mathbb{R}^{3}$ are possible angular and linear velocity vectors, and $\dot{b}\in \mathbb{R}$ stands for system's scale(or size) velocity. We call the triple 
\[\pmb{\delta}=(\pmb{v},\pmb{\omega},\dot{b})\]
system's \textit{similarity velocity}.\newline
Alternatively, we can construct a representation of the similarity group by expressing the position of the particle in absolute space $\mathbb{R}^{3}$, on the real projective space $\mathbb{R}P^{4}$ as
\[\left(\begin{array}{c}x\\ y\\z \end{array}\right)\rightarrow \left(\begin{array}{c}x\\ y\\z \\1\end{array}\right)\]
Matrices representing $Sim(3)$ on $\mathbb{R}P^{4}$ are then of the following form
\begin{equation}\label{SimRep2}
\begin{bmatrix} P_{11} &  P_{12} &  P_{13} &  t_{1} \\  P_{21} &  P_{22} &  P_{23} & t_{2}\\ P_{31} & P_{32} &  P_{33} & t_{3} \\0 & 0 & 0 & b^{-1} \end{bmatrix}
\end{equation}
A simple calculation shows indeed
\[\begin{bmatrix}P & \textbf{t} \\0 & b^{-1}\end{bmatrix}\left(\begin{array}{c}\textbf{x}\\ 1
\end{array}\right)=\left(\begin{array}{c}P\textbf{x}+\textbf{t}\\ b^{-1}\end{array}\right)\cong\left(\begin{array}{c}b(P\textbf{x}+\textbf{t})\\ 1\end{array}\right)\]
In this representation, in contrary to the previous one, any $g\in Sim(3)$ is thought of as a translation and rotation followed by a dilatation. Correspondingly, the matrix representation of the Lie-algebra $\textbf{sim}(3)$ is given by
\begin{equation}\label{simRep2}
\begin{bmatrix} 0 & \omega_{3} & -\omega_{2} & v_{1} \\ -\omega_{3} & 0 & \omega_{1} & v_{2}\\\omega_{2} & -\omega_{1} & 0 & v_{3} \\0 & 0 & 0 & -\dot{b}\end{bmatrix}
\end{equation}
The action of $Sim(3)$ on the configuration space $Q$ of a multiparticle system can consequently be given from the previous actions in a straightforward way. \newline
We use the upper left $3\times 3$ block of (\ref{SimRep1}) to construct a representation of the group
\[G_{rs}:=\frac{Sim(3)}{trans(3)}=SO(3)\times\mathbb{R}^{+}\]
comprising all the rotations and scale transformations on the center of mass configuration space $Q_{cm}$. This group has a direct product structure and acts on $\mathbb{R}^{3}$ as follows
\[ 
    \begin{bmatrix} b P_{11} & b P_{12} & b P_{13}  \\ b P_{21} & b P_{22} & b P_{23} \\b P_{31} & b P_{32} & b P_{33}  \end{bmatrix}
    \]
This action introduces a $(SO(3)\times\mathbb{R}^{+})$ fiber bundle structure on $Q_{cm}$. The Lie-algebra $\textbf{g}_{\textbf{rs}}$ of $G_{rs}$ consists of the matrices of the following form
\begin{equation}\label{Liealgebra Grs}
   \textbf{g}_{\textbf{rs}}=\textbf{so}(3)+I_{3}\dot{c}
\end{equation}
\[=\begin{bmatrix} \dot{b} & \omega_{3} & -\omega_{2}  \\ -\omega_{3} & \dot{b} & \omega_{1} \\\omega_{2} & -\omega_{1} & \dot{b}   \end{bmatrix}  \]
The letter $I_{3}$ stands for the $3\times 3$ identity matrix, and $\dot{b}\in \mathbb{R}$ for the generator of scale transformations.
One can arrive at the expression for the connection form
\[\omega=T(Q_{cm})\rightarrow \textbf{g}_{\textbf{rs}} \]
of the $G_{rs}$ fiber bundle $Q_{cm}\rightarrow S$, by modifying the connection form $\omega_{r}$ of the $SO(3)$ fiber bundle $Q_{cm}\rightarrow Q_{int}:=\frac{Q_{cm}}{SO(3)}$ in the following way
\[\omega=\omega_{r}+\omega_{s} \]
\begin{equation}\label{shapeconnection}
=R\bigg(A_{x}^{-1}(\sum_{j=1}^{N-1}\pmb{r}_{j}\times d\pmb{r}_{j})\bigg)+I_{3}D_{x}^{-1}\bigg(\sum_{j=1}^{N-1} \pmb{r}_{j}.d\pmb{r}_{j}\bigg)
\end{equation}
where $A_{x}$ stands for the moment of inertia tensor expressed in Jacobi coordinates, i.e., the following map \[A_{x}: \mathbb{R}^{3}\rightarrow \mathbb{R}^{3}\]
\begin{equation}\label{inertia}
    A_{x}(\pmb{v})=\sum_{j=1}^{N-1}\pmb{r}_{j}\times(\pmb{v}\times \pmb{r}_{j})   
\end{equation}
for any $\pmb{v}\in\mathbb{R}^{3}$ and $x\in Q_{cm}$\footnote{Strictly speaking the total collisions and linear configurations $x$ needs to be removed as the map $A$ on these regions of $Q_{cm}$ is not invertible.}. Moreover, for a velocity vector $\dot{x}\in T_{x}(Q_{cm})$
\[ \dot{x}= \begin{bmatrix}\dot{x}_{1} \\ \dot{x}_{2} \\.\\.\\.\\ \dot{x}_{3N} \end{bmatrix}=  \begin{bmatrix}\pmb{\dot{x}}_{1} \\ \pmb{\dot{x}}_{2} \\.\\.\\.\\ \pmb{\dot{x}}_{N} \end{bmatrix}\]
denote $d\pmb{x}_{i}$ as the differential 1-form defined as follows
\[ d\pmb{x}_{i}(\dot{x}):=\pmb{\dot{x}}_{i} \]
and in the same fashion, the 1-forms $d\pmb{r}_{i}$'s in (\ref{shapeconnection}) are defined for any $1\leq i \leq N-1$. The connection form $\omega_{r}$ of the $SO(3)$ fiber bundle $Q_{cm}\rightarrow Q_{int}:=\frac{Q_{cm}}{SO(3)}$, which forms the first term on the right-hand side of (\ref{shapeconnection}), is introduced in $1984$ by Alain Guichardet \cite{12}, and since then has been intensively used in the separation of rotational and vibrational motions in molecular physics and related topics, like in  \cite{1},\cite{2}, \cite{5}. The solution to the "falling cat problem" is also intimately related to the nonvanishing of the curvature of $\omega_{r}$. For a comprehensive and detailed review of this topic, reference \cite{3} is highly recommended. \newline
In the second term of the right-hand side of (\ref{shapeconnection}), we have defined the operator \[
D_{x}:\mathbb{R}\rightarrow\mathbb{R}\] as follows 
\begin{equation}\label{dilmome}
D_{x}(\dot{\pmb{\lambda}}):=\sum_{j=1}^{N-1}\pmb{r}_{j}^{2}\dot{\pmb{\lambda}}
\end{equation} 
and we call it the "\textit{dilational tensor}"\footnote{It is compatible with the definition $D=\sum_{a=1}^{N}\pmb{x}_{a}.\pmb{p}_{a}$ introduced by Barbour et al. much earlier than us.}. The letter $\dot{\pmb{\lambda}}$ stands for the rate of change of scale of the system (scale velocity)\footnote{Here, we assume all measurements are conducted using special Newtonian rods and clocks, which are isolated from the material universe and do not get affected by them in any way, or by any transformations, we perform on the material universe. Practically, such measuring instruments, of course, do not exist. However, the existence of absolute space and absolute time in the Newtonian world-view justifies their hypothetical existence in the context of this view.}
\begin{equation}
\dot{\pmb{\lambda}}:=\frac{\dot{\lambda}}{\lambda}
\end{equation}
where 
\begin{equation}\label{scvari} 
\lambda:=max\mid\textbf{x}_{i}-\textbf{x}_{j}\mid
\end{equation}
 with $i,j$ varying between $1,2,..,N$; being one choice for the system's scale variable.\newline 
We have constructed this operator in direct analogy to the moment of inertia tensor $A_{x}$. The Inertia tensor transfers an angular velocity (which can be represented as a vector in $\mathbb{R}^{3}$) to another vector in $\mathbb{R}^{3}$, which represents the total angular momentum of the whole system. In the same way, the dilational tensor $D_{x}$ transfers an expansion velocity, which in turn can be represented by a number in $\mathbb{R}$, to a measure of the total expansion of the system (dilational momentum $D$), which again can be represented by another number in $\mathbb{R}$. Since the Lie-algebra of $G_{rs}$ can be represented by the matrices (\ref{Liealgebra Grs}), one recognizes the correct structure in the connection form (\ref{shapeconnection}), for the ($SO(3)\times \mathbb{R}^{+}$) fiber bundle $Q_{cm}\rightarrow \frac{Q_{cm}}{G_{rs}}$. Taking any vector of $T_{x}(Q_{cm})$ and acting on it with this connection form, the first term yields a member of $\textbf{so}(3)$. The second term yields a number that is multiplied by the identity matrix. It results in a matrix of the above form (i.e., a member of the Lie-algebra of the bundle's structure group). So it does what it is expected to do. \newline\newline
\textbf{Curvature of the connection form $\omega_{s}$:}\newline\newline
 Last but not least, we want to investigate the curvature $C$ of the connection form
 \begin{equation}\label{fcf}
     \omega_{s}=D_{x}^{-1}\bigg(\sum_{j=1}^{N-1} \pmb{r}_{j}.d\pmb{r}_{j}\bigg)
 \end{equation}
of the $\mathbb{R}^{+}$ fiber bundle $Q_{int}\rightarrow S=\frac{Q_{int}}{Sc}$. This 1-form can equally well be expressed in the Cartesian coordinates $x_{1}$,...,$x_{3N}$ on the absolute configuration space $Q$ as follows 
 \begin{equation}\label{wsco}
     \omega_{s}=D_{x}^{-1}\big(\sum_{i=1}^{3N}m_{\lfloor\frac{i-1}{3}\rfloor +1}x_{i}dx_{i} \big)
 \end{equation}
where for every rational number $i$, the largest integer smaller than $i$ is denoted by $\lfloor i\rfloor$. \newline
Given two arbitrary horizontal vectors $v,v'\in T_{x}(Q)$ as the input of the curvature $2$-form, it is known \cite{12} that
\begin{equation}\label{cfge}
    C(v,v')=-\frac{1}{2}\omega_{s}([v,v'])
\end{equation}
where $[.,.]$ is the Lie-bracket of the extension of horizontal vectors $v$ and $v'$ to horizontal vector fields. Choosing a basis $\frac{\partial}{\partial x_{1}}$,...,$\frac{\partial}{\partial x_{3N}}$ of the tangent space $T_{x}(Q)$, Lie-bracket of the two vector fields $v=\sum_{i=1}^{3N}v_{i}\frac{\partial}{\partial x_{i}}$ and $v'=\sum_{i=1}^{3N}v_{i}'\frac{\partial}{\partial x_{i}}$ can be computed by 
\[ [v,v']=\sum_{i}\sum_{j}(v_{j}\frac{\partial v'_{i}}{\partial x_{j}}-v'_{j}\frac{\partial v_{i}}{\partial x_{j}})\frac{\partial}{\partial x_{i}} \]
As both vectors $v,v'\in T_{x}(Q)$ are horizontal, they satisfy the following conditions
\[\omega_{s}(v)=\omega_{s}(v')=0\]
Using (\ref{wsco}), the above conditions can be translated into relations    (or constraint equations) in the variables  $x_{1}$,...,$x_{3N}$,$v_{1}$,...,$v_{3N}$ which after some rearrangement of terms become as follows
\[v_{3N}=\frac{1}{m_{N}x_{3N}}\sum_{i=1}^{3N-1}m_{\lfloor\frac{i-1}{3} \rfloor +1}x_{i}v_{i} \]
\[v_{3N}'=\frac{1}{m_{N}x_{3N}}\sum_{i=1}^{3N-1}m_{\lfloor\frac{i-1}{3} \rfloor +1}x_{i}v_{i}' \]
As all other $6N-1$ variables involved are independent, for all $j=1,...,3N$, and $i=1,...,3N-1$; all the following derivatives vanish \[\frac{\partial v_{i}}{\partial x_{j}}=\frac{\partial v_{i}'}{\partial x_{j}}=0\] 
which simplify the above expression for $[v,v']$ greatly
\[ [v,v']=\sum_{j=1}^{3N}(v_{j}\frac{\partial v'_{3N}}{\partial x_{j}}-v'_{j}\frac{\partial v_{3N}}{\partial x_{j}})\frac{\partial}{\partial x_{3N}}=0 \]
In the last equality $\frac{\partial v_{3N}}{\partial x_{j}}=\frac{m_{\lfloor\frac{i-1}{3} \rfloor +1}}{m_{N}x_{3N}} v_{j}$ is used. Hence, it follows from (\ref{cfge}) that the connection form $\omega_{s}$ has vanishing curvature. This means that shape space $S=\frac{Q}{sim(3)}$ is exactly as curved as $Q_{int}:=\frac{Q}{E(3)}$. That $Q_{int}$ is curved, or in other words, that the curvature of the connection form $\omega_{r}$ is nonvanishing, had already been shown by Alain Guichardet in \cite{12}. Therefore, we can conclude from our calculations that when the flat configuration space $Q$ is quotiented by the action of the similarity group $Sim(3)$, the only step that causes curvature in the final base space is the quotienting with respect to the rotation group $SO(3)$.

\subsection{Expression of the mass metric in shape, orientation, and scale coordinates}
Decomposition of the mass metric\footnote{Or of the kinetic energy of a mechanical system} in rotational, dilational, and relational(shape) parts is the main theme of this section. The decomposition of the mass metric in rotational and internal parts on the $SO(3)$ fiber bundle $Q_{cm}\rightarrow Q_{int}=\frac{Q_{cm}}{SO(3)}$ is discussed in \cite{12},\cite{1},\cite{2}, \cite{3}. We extend the methods described there to the $G_{rs}$ fiber bundle $Q_{cm}\rightarrow\frac{Q_{cm}}{SO(3)\times Sc}$, and hence add here the scale transformations to the formalism.\newline\newline 
Associated with the group action of $SO(3)$ on $Q_{cm}$, a so-called "moving frame" \[\{\textbf{e}'_{1},\textbf{e}'_{2},\textbf{e}'_{3}\}\] of the absolute space $\mathbb{R}^{3}$ given by 
\[\textbf{e}'_{1}=g\textbf{e}_{1}, \textbf{e}'_{2}=g\textbf{e}_{2}, \textbf{e}'_{3}=g\textbf{e}_{3}\]
can be attached to the mechanical system, where \[\{\textbf{e}_{1},\textbf{e}_{2}, \textbf{e}_{3}\}\] stands for some (fixed) "space frame". The Euler angles $\phi^{1},\phi^{2},\phi^{3}=\alpha,\beta,\gamma$ are usually used to specify the orientation of the moving frame w.r.t. the space frame. Hence, any point on $Q_{cm}$ can be specified by $3N-6$ internal coordinates $q^{i}$ relative to the body frame and the three Euler angles $\phi^{a}$. The components of a vector in $\mathbb{R}^{3}$ with respect to the body frame will be denoted by a subscript "b". The relation between the expression of a vector in the space frame and the body frame is known to be as follows
\[\pmb{v}_{b}=g^{-1}\pmb{v}_{s} \]
where $g$ is the rotation that takes the space frame to the body frame. For notational simplicity, from now on, we drop the subscript $s$ whenever we mean the expression of vectors w.r.t. the space frame. \newline
There are two different kinds of vector fields defined on the group $SO(3)$ 
denoted by $L_{a}$ and $J_{a}$, for $a=1,2,3$. The first set of vector fields $L_{a}$ coincides with the direction of movement along the fiber (which is itself part of $T(Q_{cm})$) if we rotate the system in $\mathbb{R}^{3}$ around the $a$'th \emph{axis of body frame} ($\textbf{e}'_{a}=g\textbf{e}_{a}$), without changing its shape. Their dual one-forms are denoted by $\theta^{a}$. The second set of vector fields $J_{a}$, coincides with the direction of movement along the fiber if we rotate the system in $\mathbb{R}^{3}$ around the $a$'th \emph{axis of space frame} $\textbf{e}_{a}$, without changing its shape. Their dual one-forms are denoted by $\psi^{a}$.  
\newline
Denote the components of the connection form $\omega_{r}$ with respect to the laboratory frame $\textbf{e}_{a}$ and the moving frame $\textbf{e}'_{a}$ as follows 
\[\omega=\sum_{a=1}^{3}R(\textbf{e}_{a})\omega^{a}=\sum_{a=1}^{3}R(\textbf{e}'_{a})\omega'^{a}\] 
where $\omega^{a}:=\omega .R(\textbf{e}_{a})$, $\omega'^{a}:=\omega .R(\textbf{e}'_{a})$. \newline
Consider a local trivialization of the center of mass configuration space\footnote{Technically speaking, we mean the non-singular stratum of configuration space.} \[Q_{cm}\cong \mathbb{R}^{3N-3}\cong I\times SO(3)\] 
with $I\subset Q_{int}:=\frac{Q_{cm}}{SO(3)}$. By writing out the connection form $\omega_{r}$ in the coordinates $(q,g)$ on this local trivialization of $Q_{cm}$ one can give a local expression of $\omega^{a}$ as follows (\cite{1}, \cite{2})
\begin{equation}\label{connlab}
\omega^{a}=\psi^{a}+\sum_{i=1}^{3N-6}\wedge^{a}_{i}dq^{i}
\end{equation}
\begin{equation}\label{connmov}
\omega'^{a}=\theta^{a}+\sum_{i=1}^{3N-6}\wedge'^{a}_{i}dq^{i}
\end{equation}
where, as before, $\psi^{a}$ and $\theta^{a}$ are respectively the three right and left invariant one-forms on $SO(3)$ defined through 
\begin{equation}\label{riof}
  dgg^{-1}=:\sum_{a=1}^{3}\psi^{a}R(\textbf{e}_{a})  
\end{equation}
\begin{equation}\label{liof}
    g^{-1}dg=:\sum_{a=1}^{3}\theta^{a}R(\textbf{e}_{a})
\end{equation}
Because changing the shape of a mechanical system causes extra total rotation and makes the direction of the movement of the configuration point deviate from $L_{a}$ or $J_{a}$, depending on how fast and in which way the system's shape is changing, one needs the second terms in (\ref{connlab}) and (\ref{connmov}).\newline\newline

Consider the $G_{rs}$ fiber bundle $Q_{cm}\rightarrow S$, and take the following local coordinates 
\[(s,g,\lambda,\dot{s},\dot{g},\dot{\pmb{\lambda}})\] on $T\big(\pi^{-1}(U)\big)\subset T(Q_{cm})$ with $U\subset S$. They are adopted coordinates to the bundle's projection map $\pi:Q_{cm}\rightarrow S$. So \[(s,g,\lambda)\in \pi^{-1}(U)\] and \[(\dot{s},\dot{g},\dot{\pmb{\lambda}})\in T_{\sigma(s)}(\pi^{-1}(U))\] 
Here $s=(s^{\alpha})$ with $\alpha\in [1,...,3N-7]$, are for instance the $3N-7$ independent angles between the $N-1$ Jacobi vectors $\textbf{r}_{i}$, and $\dot{\pmb{\lambda}}:=\frac{\dot{\lambda}}{\lambda}$ is the scale velocity, and the map 
\[\sigma:U\rightarrow  Q_{\textit{cm}}\] 
defines a local section(or lift) from shape space to the center of mass configuration space. Then, any point $x\in \pi^{-1}(U)\subset Q_{cm}$ can be expressed as 
\begin{equation}
 x=\lambda g\sigma(s)=\big(\lambda g\pmb{\sigma}_{1}(s),...,\lambda g\pmb{\sigma}_{N-1}(s)\big)
\end{equation}
with $g\in SO(3)$, $\lambda\in Sc$ and $s\in U\subset S$. Note that \[\sigma(s)= \left(\begin{array}{c}\pmb{\sigma}_{1}(s)\\...\\ \pmb{\sigma}_{N-1}(s) \end{array}\right)=\left(\begin{array}{c}\sum_{a=1}^{3}C_{1}^{a}\textbf{e}_{a}\\...\\  \sum_{a=1}^{3}C_{N-1}^{a}\textbf{e}_{a}\end{array}\right)\] 
basically describes a way to put the multi-particle system with shape $s\in U$ in the absolute space $\mathbb{R}^{3}$. This allocation is achieved by choosing the point $x\in Q_{cm}$ on the $G_{rs}$ fiber above $s$, to which the shape $s$ is meant to be lifted as follows
\[\pmb{r}_{3(i-1)+a}:=C_{i}^{a} \]
So specifying a section $\sigma$, comes down to the specification of a set of $3(N-1)$ real valued functions on shape space, i.e. \[C_{i}^{a}:S\rightarrow \mathbb{R}\] 
The \textbf{horizontal lift}\footnote{ $\omega_{\lambda g\sigma(s)}\big((\frac{\partial}{\partial s^{\alpha_{0}}})^{*}\big)=0$ and $\pi_{*}\big((\frac{\partial}{\partial s^{\alpha_{0}}})^{*}\big)=\frac{\partial}{\partial s^{\alpha_{0}}}$ are the defining criteria of horizontal lift of a local vector field $\frac{\partial}{\partial s^{\alpha_{0}}}$ on $U\subset S$.} $(\frac{\partial}{\partial s^{\alpha_{0}}})^{*}\in T_{s_{0}}(S)$, of a local vector field $\frac{\partial}{\partial s^{\alpha_{0}}}$ on $U\subset S$ to a point $x\in Q_{cm}$ is given by 
\begin{equation}\label{hlfs}
    (\frac{\partial}{\partial s^{\alpha}})^{*}=\frac{\partial}{\partial s^{\alpha}}-D_{r}^{-1}\big(\sum_{j=1}^{N-1}\pmb{r}_{j}.   \frac{d\pmb{r}_{j}}{ds^{\alpha}} \big)\frac{\partial }{\partial \lambda } 
\end{equation}
\[-\sum_{a=1}^{3}\beta_{\alpha}^{a}(s_{0})L_{a} \]
with (see \cite{2})
 \[\beta^{a}_{\alpha}:=\big<A_{\sigma(s_{0}) }^{-1}\big(\sum_{i=1}^{N-1}\pmb{r}_{i}\times\frac{\partial \pmb{r}_{i}}{\partial s^{\alpha}}\big)\mid e'_{a}\big>\] 
Expression (\ref{hlfs}) can be derived by requiring $\omega((\frac{\partial}{\partial s^{\alpha_{0}}})^{*})=0$, which is of course how a horizontal lift w.r.t. a connection should be.\newline
  A local basis of one-forms and vector fields on $\pi^{-1}(U)\cong U\times G_{rs}$ can be formed from 
\[ds^{\alpha},\omega^{\prime a}_{r},\omega_{s}\] and 
\[ (\frac{\partial}{\partial s^{\alpha}{}})^{*},L_{a},\frac{\partial}{\partial \lambda}\] 
  respectively, with $\omega_{r}$ and $\omega_{s}$ denoting the rotational and the dilational part of the $Sim(3)$-connection form (\ref{shapeconnection}) which we introduced before.
They are in accordance with the decomposition $T_{x}(Q_{cm})=V_{x}\oplus H_{x}$. Technically speaking, we have to use $\pi^{*}ds^{\alpha}$, the pullback of $ds^{\alpha}$ under the bundle's projection map, but for the sake of notational simplicity, we still used $ds^{\alpha}$. 
\newline \newline 
Having now the connection form $\omega_{r}$, i.e., first part of (\ref{shapeconnection}), in mind, one can introduce a \textbf{so}(3)-valued variable (\cite{1},\cite{2}) as follows
\begin{equation}
\Pi = \epsilon + \sum_{\alpha=1}^{3N-6}\wedge_{\alpha}(x)\dot{q}^{\alpha}
\end{equation}
where 
\[\epsilon=g^{-1}\dot{g}\] and
\[\wedge_{\alpha}(x)=\sum_{a=1}^{3}\wedge_{\alpha}^{a}(x)R(\textbf{e}_{a})\]
The vectors associated with $\Pi$ and $\epsilon$ will be denoted by $\Omega'$ and $\Omega$, respectively, i.e., \[R(\Omega')=\Pi\] and \[R(\Omega)=\epsilon\] 
Thus, the tuple \[(s,g,\lambda,\dot{s},\Omega',\dot{\pmb{\lambda}})\] constitutes a local (quasi)coordinate system on $T(\pi^{-1}(U))$. Bear in mind that $\Omega'$ denotes the angular velocity of the system in the body frame, and hence the angular momentum of the system in the space frame would become $\textbf{L}=gA_{\sigma(s)}\Omega'$. Therefore, the angular momentum vector expressed in the body frame becomes $A_{\sigma(s)}\Omega'$. As usual, $g$ stands for the rotation, which brings the space frame to the body frame. \newline
According to the orthogonal decomposition \[T_{x}(Q_{cm})=V^{dilational}_{x}\oplus V^{rotational}_{x}\oplus H_{x}\] 
the mass metric $\textbf{M}$ (\ref{KEN}) in coordinates $(s,\lambda,\alpha,\beta,\gamma)$ can be expressed in terms of the following line element:
\begin{equation}\label{mminnewcor}
dl^{2}=\sum_{\alpha,\beta=1}^{3N-7}N_{\alpha\beta}ds^{\alpha}ds^{\beta}+\sum_{i=1}^{N-1}\mid\frac{\pmb{r}_{i}}{\lambda}\mid^{2} (d\lambda)^{2}
\end{equation}\[ +\sum_{a,b=1}^{3}A_{ab}\omega_{r}'^{a}\omega_{r}'^{b}\]
where
\[\left\{
  \begin{array}{lr}
   N_{\alpha\beta}:=ds^{2}((\frac{\partial}{\partial s^{\alpha}})^{*},(\frac{\partial}{\partial s^{\beta}})^{*})\\
    A_{ab}:=dl^{2}(J_{a},J_{b})=\textbf{e}_{a}.A_{\sigma(s)}(\textbf{e}_{b})
  \end{array}
\right.
\]
and $\pmb{r}_{i}$'s are as before the Jacobi vectors of the system, and hence are unique functions of the $s_{\alpha}$'s and $\lambda$.
Remember that $\omega_{r}^{a}(J_{b})=\psi^{a}(J_{b})=\delta^{a}_{b}$, and that $N_{\alpha\beta}$ defines a Riemannian metric on the shape space $\frac{Q_{cm}}{G_{rs}}$, and $A$ is the inertia tensor. The $\omega_{r}'^{a}$'s are components of the connection form of rotations $\omega_{r}$(first part of \ref{shapeconnection}) in body frame (\ref{connmov}). $\omega'_{r}$ and $\Omega'$ are dual to each other, i.e. $\omega'(\Omega')=1$.\newline
By setting the instantaneous unit of length equal to the system's scale variable $\lambda$, one gets the following expression for the metric
\begin{equation}\label{metric siminv}
ds^{2}=\sum_{\alpha,\beta=1}^{3N-7}N_{\alpha\beta}ds^{\alpha}ds^{\beta}
+\sum_{i=1}^{N-1}\mid \pmb{r}_{i}\mid^{2} (d\lambda)^{2}\end{equation}\[+\sum_{a,b=1}^{3}A_{ab}\omega'^{a}\omega'^{b}\]
where now all the $\pmb{r}_{i}$'s and $A$ are expressed in the internal (expanding or contracting or stationary\footnote{Compared with absolute length unit.}) length unit. It is worth mentioning that as the independent angles $s_{i}$ and the scale coordinate bring the metric tensor in a diagonal form, they form an orthogonal coordinate system on $Q_{int}=\frac{Q_{cm}}{SO(3)}$.

\section{Reduced equations of motion on shape space}
For the reduction of modified Newtonian mechanics with respect to the group of similarity transformations $Sim(3)$, we use the geometric setting explained in the last section. One of the reasons why the reduction w.r.t. the similarity group is not studied as extensively as the Euclidean group $E(3)$ is that the potential function of a classical system defined on the absolute configuration space $Q$, while rotation and translation invariant, is not scale invariant (take the Newtonian gravitational potential, or the Columb potential as examples). However, as mentioned before, scale transformations become an additional symmetry in modified classical physics due to its scale invariant potential function. This additional symmetry immediately allows the shape degrees of freedom to have an autonomous evolution, entirely decoupled from the system's transnational, orientational, and scale degrees of freedom. \newline
In this section, we seek the equations of motion of an $N$-particle system in shape, orientation, and scale coordinates and their velocities. For this purpose, the Lagrangian of the system must first be expressed in terms of the new coordinates and velocities, and then the equations of motion can be derived. Since the angular velocities used to quantify the rate of rotation of a system are not derivatives of the three Euler angles (or any other variables), the Lagrange or Euler-Lagrange equations of motion cannot be used. In such cases, the Boltzmann-Hamel equations of motion must be used. The constraint of constant angular velocity is a nonholonomic constraint for the motion of mechanical systems, just like rolling without slipping a rigid body on a surface. The dynamics of nonholonomic systems is a well-studied topic, and there are comprehensive monographs dealing with all aspects of such systems, e.g., \cite{9}, \cite{99}. The reduced equations of motion with respect to symmetry groups are also known as the Lagrange-Poincare equations\cite{MarsdenRatiuScheurle},\cite{MestdagCrampin}.\newline
 In the first subsection, following \cite{9}, we will briefly review the formulation of mechanics in quasi-coordinates and quasi-velocities. We then derive the equations of motion of modified classical systems on shape space by extending the results of \cite{1} and \cite{2}\footnote{which are restricted to the reduction w.r.t. the Euclidean group} to the scale transformations.
\subsection{Equations of motion in quasi-velocities}
The generalized coordinates are the set of coordinates defining the degrees of freedom of a system. For instance, for a rigid body moving in $\mathbb{R}^{3}$, there are six generalized coordinates (three specifying the position of the body and three the orientation of it), i.e., \[{q}=[q_{1},...,q_{6}]:=[ x, y, z, \alpha, \beta, \gamma ]\] 
The generalized speeds are the derivatives of the generalized coordinates ${\dot{q}}=[\dot{x}, \dot{y}, \dot{z}, \dot{\alpha}, \dot{\beta}, \dot{\gamma}]$. These coordinates $q_{k}$ can be called \textit{true coordinates}, in a sense that if the velocities $\dot{q}_{k}$ are known functions of time, integration with respect to time determines their respective coordinates, and hence the state of the system. On the other hand, one may define generalized speeds, which are not integrable\footnote{which cannot be written as the time derivative of any coordinates; for instance, they are defined as linear combinations of the time derivatives of generalized coordinates.}. Such generalized speeds are called \textit{quasi-velocities}. As the most famous example of quasi-velocities, one can mention angular velocity components of a rigid body, which are linear combinations of derivatives of Euler angles. However, they are themselves not time derivatives of any coordinates. Quasi-velocities were first introduced to derive the so-called Boltzmann-Hamel equations of motion, which we will discuss shortly below.\newline 
In analyzing non-holonomic systems, quasi-velocity formulation casts the dynamical equations of motion in a form requiring fewer equations. For a system possessing $n$ degrees of freedom with $m$ non-holonomic constraints, the usage of Lagrangian formalism leads to $2n+m$ equations of motion ($2n$ equations for the system's state and $m$ algebraic relations that must be solved for the multipliers). However, If quasi-velocity formalism is used, the same problem can be described by a system of $2n-m$ degrees of freedom (see, for instance, \cite{6},\cite{7}). \newline\newline
Equations of motion for classical mechanics in true coordinates are the known Lagrange equations 
\begin{equation}\label{rle}
    \frac{d}{dt}\frac{\partial K}{\partial \dot{q}_{i}}-\frac{\partial K}{\partial q_{i}}=F_{i}
\end{equation}
with $K$ being the system's kinetic energy, and $F_{i}$ being the generalized force associated with the generalized coordinate $q_{i}$, for $i\in[1,n]$. \newline
Consider now the usage of the quasi-velocities $Y_{i}$, which are defined as $n$ independent linear combinations of the $\dot{q}_{k}$'s, i.e., \[Y_{i}:=\alpha_{i1}\dot{q}_{1}+\alpha_{i2}\dot{q}_{2}+...+\alpha_{in}\dot{q}_{n}=\sum_{r=1}^{n}\alpha_{ir}\dot{q}_{r} \]
with $\alpha_{ir}$ being known functions of the generalized coordinates $q_{k}$.
Constructing a $n\times n$ matrix $\alpha$ from $\alpha_{ij}$'s, one can write the definition of quasi-velocities $Y_{k}$'s more compactly as follows
\[ \begin{bmatrix}Y_{1} \\ Y_{2} \\.\\.\\.\\ Y_{n} \end{bmatrix}=\alpha  \begin{bmatrix}\dot{q}_{1} \\ \dot{q}_{2} \\.\\.\\.\\ \dot{q}_{n} \end{bmatrix}\]
Having the above relations between the quasi-velocities $Y_{i}$'s and the  generalized (true) velocities $\dot{q}_{j}$'s in mind, one can define a set of differential forms $d y_{k}$ as follows
\[d y_{k}:=\sum_{r=1}^{n}\alpha_{rk}dq_{r} \]
The above equations cannot always be integrated to obtain the variable $y_{k}$. In such cases, the differential form $dy_{k}$ is naturally called "nonintegrable" and cannot be thought of as differential of some configuration variable $y_{k}$. The quantities $d y_{k}$ are called \textit{differentials of quasi-coordinates}, with some abuse of words because they are not really differentials, and the variables $y_{k}$ are undefined.\newline
If the quasi-velocities are known, the true velocities can be calculated using \[\dot{q}_{k}=\beta_{kl}Y_{l}\] where $\alpha_{sk}\beta_{kl}=\delta_{sl}$. Here $\delta_{sl}$ is the Kronecker Delta. It is easy to check that
\[ \frac{\partial\dot{q}_{k}}{\partial Y_{l}}=\frac{\partial q_{k}}{\partial y_{l}}=b_{kl}\]
For a function $f(q_{1},...,q_{n},t)$, with the partial derivative with respect to a quasi-coordinate $y_{l}$ one means the following \[\frac{\partial f}{\partial y_{l}}:=\frac{\partial f}{\partial q_{k}}\frac{\partial q_{k}}{\partial y_{l}}=\frac{\partial f}{\partial q_{k}}b_{kl} \]
The kinetic energy $K(q_{1},...,q_{n},\dot{q}_{1},...,\dot{q}_{n})$ can be expressed in the new variables, i.e., $K'(q_{1},...,q_{n},Y_{1},...,Y_{n})$, where the prime here indicates just the difference in the variables of the function.
The equations of motion expressed in these new coordinates $q_{1},...,q_{n},Y_{1},...,Y_{n}$ are known as \textit{Boltzmann-Hamel} equations and become of the following form (see \cite{9}) 
 \begin{equation}\label{emqq}
     \frac{d}{dt}\frac{K'}{\partial Y_{k}}-\frac{\partial K'}{\partial y_{k}}+\gamma_{kij}\frac{\partial K'}{\partial Y_{i}}Y_{j}=F'_{k}
  \end{equation}
for $k=1,2,...,N$ where \[F'_{k}=\sum_{s=1}^{n}F_{i}b_{ik}\] are the generalized forces corresponding to the virtual displacements $\delta y_{k}$, and 
\[\gamma_{kij}:=b_{sk}b_{lj}\big(\frac{\alpha_{ij}}{q_{l}}-\frac{\partial \alpha_{il}}{\partial q_{s}} \big) \]
As already mentioned, the partial derivative with respect to the quasi-coordinate appearing in the second term of (\ref{emqq}) should be understood as follows \[ \frac{\partial K'}{\partial y_{k}}:=\frac{\partial f}{\partial q_{l}}\frac{\partial q_{l}}{\partial y_{k}}=\frac{\partial f}{\partial q_{l}}b_{lk}\] 
The $n$ equations of (\ref{emqq}) are equations of motion in quasi-coordinates. If $y_{1}$,...,$y_{n}$ are true coordinates the coefficients $\gamma_{kij}$ all vanish, and the Boltzmann-Hamel equations (\ref{emqq}) take back the form of Lagrange equations (\ref{rle}).

\subsection{Lagrangian of MNT in quasi-velocities}
After expressing the kinetic metric in new coordinates (\ref{metric siminv}) and the discussion about the potential function $V$ in Section (II.v) and this section's introduction, respectively, we are in a position to write down the similarity invariant Lagrangian of the modified Newtonian mechanics\footnote{For the non-singular configurations.} as follows
\begin{equation}\label{siminla}
\mathscr{L}=\frac{1}{2}\sum_{\alpha,\beta=1}^{3N-7}N_{\alpha\beta}\dot{s}^{\alpha}\dot{s}^{\beta}+\frac{1}{2}\sum_{i=1}^{N-1}(\pmb{r}_{i}\dot{\pmb{\lambda}})^{2}
\end{equation} 
\[ +\frac{1}{2}\sum_{a,b}A_{ab}\Omega'^{a}\Omega'^{b}-V(s)\]
Here $V$ is a similarity invariant potential function, hence depends only on the $3N-7$ coordinates $s_{i}$. Notice that our previous expression for dilational momentum (\ref{dilmome}) is also derivable from the above Lagrangian as follows \[D=\frac{\partial L}{\partial \pmb{\dot{\lambda}}}=\sum\mid \pmb{r}_{j}\mid^{2}\pmb{\dot{\lambda}}\]  As the lagrangian (\ref{siminla}) is scale independent, $D$ is a constant of motion. Remember that the units in which $D$ is constant are all internal units. In absolute(external) units, if at a given instant of time $D>0$, then $D$ will monotonically increase with (absolute)time. Accelerated expansion of a modified Newtonian universe, discussed in \cite{Me4}, is a consequence of the mentioned property of dilational momentum $D$ in MNT.  \newline\newline
Modified Newtonian Theory has a similarity invariant Lagrangian $\mathscr{L}(s,g,\lambda,\dot{s},\Omega',\dot{\pmb{\lambda}})$, i.e.,
\begin{equation}\label{il}
\mathscr{L}(s,hg,b\lambda,\dot{s},\Omega',\dot{\pmb{\lambda}})=\mathscr{L}(s,g,\lambda,\dot{s},\Omega',\dot{\pmb{\lambda}})
\end{equation}
$\forall h\in SO(3)$, and $\forall b \in \mathbb{R}^{+}$. Note that $\Omega'$ is left $SO(3)$ invariant. Such a function $\mathscr{L}$ on $T(Q_{cm})$ reduces naturally to a function $\mathscr{L}^{*}(s,\dot{s},\Omega',\dot{\pmb{\lambda}})$ on $\frac{T(Q)}{Sim(3)}$. The similarity invariance of the Lagrangian in modified classical mechanics is a consequence of the similarity invariance of kinetic and potential energies, as already explained based on the principle of relationalism. It is important to remember that the units (of time and spatial distance) in which the Lagrangian has the property (\ref{il}) are internal units.    

\subsection{Reduced Euler-Lagrange equations of motion}
In this section, finally, we discuss  the equations of motion in the non-holonomic frame $(s,g,\lambda,\dot{s},\Omega',\dot{\pmb{\lambda}})$ on $T(Q_{cm})$ introduced previously.\newline\newline
Let $x^{\lambda}$, $\lambda=1,...,3N-3$ be a local coordinate system on $W\subset Q_{cm}$. From this coordinate system, one can derive a basis for the vector fields, i.e., $\frac{\partial}{\partial x^{\lambda}}$, and a basis for the 1-forms, i.e., $dx^{\lambda}$ on $T(Q_{cm})$. Let $Z_{\lambda}$ and $Z^{\lambda}$ be another local basis of the vector fields and 1-forms (dual to each other) on $W$. The later vector fields and one-forms are related to the former ones\footnote{Which were derived from the coordinate system $x^{\lambda}$} by
\[Z_{\lambda}=\sum_{\mu=1}^{3N-3}E^{\mu}_{\lambda}\frac{\partial}{\partial x^{\mu}}\]
\[Z^{\lambda}=\sum_{\mu=1}^{3N-3}E^{\prime\lambda}_{\mu}dx^{\mu}\]
where their duality requires $\sum_{\lambda}E'^{\mu}_{\lambda}E^{\lambda}_{\nu}=\delta^{\mu}_{\nu}$. If the above relations for $Z^{\lambda}$'s are integrable, there exists true coordinates $z^{\lambda}$ on $Q_{cm}$, for which $Z_{\lambda}=\frac{\partial}{\partial z^{\lambda}}$, and $Z^{\lambda}=dz^{\lambda}$. Otherwise, the $Z_{\lambda}$'s form an non-holonomic basis of $T(Q_{cm})$, and the $z^{\lambda}$'s become the corresponding quasi-coordinates on $Q_{cm}$. Differentiation of $Z^{\lambda}$ leads to \cite{2}
\[dZ^{\lambda}=\sum_{\sigma<\kappa}\gamma^{\lambda}_{\sigma\kappa}Z^{\kappa}\wedge Z^{\sigma} \]
with
\[\gamma^{\lambda}_{\sigma\kappa}:=\sum_{\mu\nu}\bigg(\frac{\partial E'^{\lambda}_{\mu}}{\partial x^{\nu}}- \frac{\partial E'^{\lambda}_{\nu}}{\partial x^{\mu}}\bigg)E_{\sigma}^{\mu}E_{\kappa}^{\nu}\]
In the expression of the Lagrangian function $\mathscr{L}(x,\dot{x})$, one can replace the coordinate velocities $\dot{x}^{\lambda}$ with the new set of velocities\footnote{Which typically are quasi-velocities} 
\[\dot{z}^{\lambda}=\sum_{\mu}E'^{\lambda}_{\mu}(x)\dot{x}^{\mu}\]
In these new variables the Lagrangian is denoted by $\mathscr{L}'$, i.e.,
\[\mathscr{L}'(x,\dot{z})=\mathscr{L}(x,\dot{x})\]
As reviewed shortly in Section (III.i), the Boltzmann-Hamel equations of motion in terms of $(x^{\lambda},\dot{z}^{\lambda})$ takes the following form
\begin{equation}\label{ELquasi}
\frac{d}{dt}\big(\frac{\partial\mathscr{L}'}{\partial\dot{z}^{\sigma}}\big)-Z_{\sigma}\mathscr{L}'+\sum_{\mu,\kappa}\gamma^{\mu}_{\sigma\kappa}\frac{\partial \mathscr{L}'}{\partial\dot{z}^{\mu}}\dot{z}^{\kappa}=0 
\end{equation}
for $\sigma=1,...,3N-3$. A derivation of the reduced equations of motion on $Q_{int}=\frac{Q_{cm}}{SO(3)}$ is explained in \cite{2}. In the rest of this section, we will present an extension of this work to find the reduced Lagrangian equations of motion on shape space $S=\frac{Q_{cm}}{SO(3)\times Sc}$.\newline
So we consider $Q_{cm}$ as a $SO(3)\times Sc$ fiber bundle and make a coordinate transformation on $Q_{cm}$ from the Euler angles, and Jacobi coordinates to the Euler angles, scale, and the shape coordinates as follows
\[(\textbf{r}_{1},\textbf{r}_{2},...,\textbf{r}_{N-1},\alpha,\beta,\gamma)\rightarrow (s_{1},...,s_{3N-7},\lambda, \alpha,\beta,\gamma)        \]
A basis of $1$-forms on $Q_{cm}$ in the new coordinates is given as follows 
\[Z^{a}:=\omega_{r}^{a}\]
\[Z^{4}:=\omega_{s}\]
\[Z^{4+i}:=ds^{i}\]
with $a=1,2,3$ and $i=1,2,...,3N-7$. The $\omega_{r}^{a}$'s and $\omega_{s}$ can be read from the new connection form (\ref{shapeconnection}). In particular, the $\omega_{r}^{a}$'s are components of the rotations connection form (\ref{connlab}) form (w.r.t. the fixed space frame). Their dual vector fields are as follows
\[Z_{a}=J_{a}\]
\[Z_{4}=\frac{\partial }{\partial\lambda }  \]
\[Z_{4+i}=\partial^{*}_{4+i}=(\frac{\partial}{\partial s^{i}})^{*}\]
where $(\frac{\partial}{\partial s^{i}})^{*}$ denotes the horizontal lift (\ref{hlfs}) of a vector $\frac{\partial}{\partial s^{\alpha}}\in T_{s_{0}}(S)$ at the point $s_{0}$ of shape space $S$, to the point $\sigma(s_{0})\in Q_{cm}$ along section (or a lifting map) $\sigma:S\rightarrow Q_{cm}$.\newline
Taking the exterior derivatives of $Z^{\lambda}$ leads to the factors $\gamma^{\lambda}_{\sigma\kappa}$. They become
\[\gamma^{a}_{bc}=-\epsilon_{bca} \]
\[\gamma^{a}_{4+i,4+j}=-\pmb{k}^{a}_{ij}\]
 with $\pmb{k}^{a}_{ij}$ 
\[\pmb{k}^{c}_{ij}=\frac{\partial\beta^{c}_{j}}{\partial s^{i}}-\frac{\partial\beta^{v}_{i}}{\partial s^{j}}-\sum_{a,b=1}^{3}\epsilon_{abc}\beta^{a}_{i}\beta^{b}_{j} \]
and all other $\gamma^{\lambda}_{\sigma\kappa}$ vanishing.\newline
$\pmb{k}^{c}_{ij}$ are the components of the curvature tensor of shape space
\[\pmb{k}^{c}=d\omega^{c}-\sum_{a<b}^{3}\epsilon_{abc}\omega^{a}\wedge \omega^{b}=\sum_{i<j}^{3N-7}\pmb{k}^{c}_{ij}ds^{i}\wedge ds^{j} \] 
for the connection form $\omega$ of $(SO(3)\times Sc)$ fiber bundle given by (\ref{shapeconnection}). 
As calculated at the end of Section (II.iii), because the connection form $\omega_{s}$ on $Q_{int}$ considered as $Sc$ fiber bundle is flat (\ref{fcf}), the above curvature tensor on $S$ becomes the same as the curvature tensor on $Q_{int}=\frac{Q_{cm}}{SO(3)}$ derived in \cite{2}.  \newline
Note that besides the well-known interconnection of changes in Euler angels, which manifest themselves in the structure constants $\gamma^{a}_{bc}=-\epsilon_{abc}$, the only non-vanishing couplings are the coupling of shape variables $s^{i}$, to the orientational variables (Euler angles). Intuitively, as the scale of a mechanical system can be changed without any resultant changes in either the total orientation or shape of the system, one expects the vanishing of corresponding $\gamma$ factors, which is just another expression of the flatness of the $\omega_{s}$ given by (\ref{fcf}).  \newline \newline
Consider the coordinates \[(\alpha, \beta, \gamma, \lambda, s^{i};\Omega^{1},\Omega^{2},\Omega^{3},\dot{\pmb{\lambda}},\dot{s}^{i})\] on $T(Q_{cm})$, where the quasi-velocities $\Omega^{a}$'s are defined using (\ref{connlab}), i.e. 
\[\Omega^{a}:=\omega_{r}^{a}(\frac{d}{dt})=\psi^{a}_{t}+\sum_{i}\beta^{a}_{i}\dot{s}^{i}=\psi^{a}(\frac{d}{dt})+\sum_{i}\beta^{a}_{i}\dot{s}^{i}\]
These are the components of angular velocity with respect to the fixed space frame.\newline 
As seen before (\ref{siminla}), the Lagrangian describing a mechanical $N$-particle system expressed in the above coordinates becomes as follows 
\[ \mathscr{L}=\frac{1}{2}\sum_{\alpha,\beta=1}^{3N-7}\textbf{N}_{\alpha\beta}\dot{s}^{\alpha}\dot{s}^{\beta}+\frac{1}{2}\sum_{i=1}^{N-1}(\pmb{r}_{i}\dot{\pmb{\lambda}})^{2}\]\[+\frac{1}{2}\sum_{a,b}A_{ab}\Omega^{a}\Omega^{b}-V(s)\]
The Boltzmann-Hamel equations of motion (\ref{ELquasi}) in this new coordinate system then take the following form:
\begin{equation}\label{shad} 
\frac{d}{dt}\big(\frac{\partial \mathscr{L}}{\partial \dot{s}^{i}}\big)-(\frac{\partial}{\partial s^{i}})^{*}\mathscr{L}-\sum_{a}\sum_{j}\pmb{k}^{a}_{ij}\frac{\partial \mathscr{L}}{\partial \Omega^{a}}\dot{s}^{j}=0
\end{equation} 

\begin{equation}\label{rotd} 
\frac{d}{dt}\big(\frac{\partial \mathscr{L}}{\partial \Omega^{a}}\big)-J_{a}\mathscr{L}-\sum_{b,c}\epsilon_{acb}\frac{\partial \mathscr{L}}{\partial \Omega^{b}}\Omega^{c}=0
\end{equation}
\begin{equation}\label{scad} 
\frac{d}{dt}\big(\frac{\partial\mathscr{L}}{\partial\pmb{\dot{\lambda}}}\big)-\frac{\partial\mathscr{L}}{\partial \lambda}=0
\end{equation}
Specifically, (\ref{scad}) expresses the conservation of the dilational momentum, and it is explained in \cite{Me4} how this new conservation law leads to an observed accelerated expansion of the whole system (universe). The $3N-7$ equations (\ref{shad}) constitute the \textit{reduced equations of motion of MNT} on shape space. \newline \newline
As we will see explicitly in the next section, in the most general case, in the above equations of motion of the modified Newtonian theory on shape space, both $D$ and $L$ appear as two single numbers. If one wishes, one can interpret these numbers as the conserved total angular momentum and the conserved total dilatational momentum of the universe in absolute space\footnote{In the language of the DGZ-approach, this corresponds to a particular choice of gauge.}. However, this interpretation is meaningless in the Leibnizian worldview. There, one must consider these two constants as part of the fundamental law of motion (of the modified Newtonian theory) on shape space. In the future, an explanation or argumentation for their values could be found based on the shape space.
According to the (modified) Newtonian theory, due to the absence of the Coriolis and centrifugal forces in the cosmological frame of reference\footnote{Built from distant galaxies of the universe.}, the total $L$ of our universe is zero and therefore does not appear in the evolution of the shape space of the universe. In contrast, there is a $D$ (or $\pmb{\dot{\lambda}}$) involved in the equations of motion of the shape of the universe. Again, as mentioned above, a true relationalist should view this constant number $D$ appearing in the equation of motion primarily as part of the law of motion on shape space. In contrast, an absolutist sees the origin of $D$ in the dilatational momentum of our universe in absolute space. \newline
The Couchy data for the modified Newtonian theory consists of the shape, the shape velocity, the total $L$, and the total $D$. By absorbing the last two in the law of motion, the specification of a point and a velocity on shape space would be sufficient to determine an entire history (solution).

\section{Three body system}
Consider three particles located at the positions \[\pmb{x}_{1}=(x_{1},y_{1},z_{1})\] \[\pmb{x}_{2}=(x_{2},y_{2},z_{2})\] \[\pmb{x}_{3}=(x_{3},y_{3},z_{3})\]
Two following Jacobi vectors can characterize the configuration of this system in the center of mass frame
\[\pmb{r}_{1}=(\frac{1}{m_{1}}+\frac{1}{m_{2}})^{-1/2}(\pmb{x}_{2}-\pmb{x}_{1})\]
\[\pmb{r}_{2}=(\frac{1}{m_{1}+m_{2}}+\frac{1}{m_{3}})^{-1/2}(\pmb{x}_{3}-\frac{m_{1}\pmb{x}_{1}+m_{2}\pmb{x}_{2}}{m_{1}+m_{2}}) \]
As shape variables, we introduce the two angles formed by the inter-particle vectors, i.e.,
\[s_{1}:=cos^{-1}(\frac{(\pmb{x}_{2}-\pmb{x}_{1}).(\pmb{x}_{3}-\pmb{x}_{1})}{\mid\pmb{x}_{2}-\pmb{x}_{1}\mid \mid \pmb{x}_{3}-\pmb{x}_{1} \mid})\]
\[s_{2}:=cos^{-1}(\frac{(\pmb{x}_{3}-\pmb{x}_{2}).(\pmb{x}_{1}-\pmb{x}_{2})}{\mid\pmb{x}_{3}-\pmb{x}_{2}\mid \mid \pmb{x}_{1}-\pmb{x}_{2} \mid})\]
and the scale variable of the system is chosen like in (\ref{scvari}) as follows
\[\lambda:=max\mid\pmb{x}_{i}-\pmb{x}_{j}\mid\]
The system's rotational degrees of freedom can be coordinatized by the three Euler angles $\alpha,\beta,\gamma$, which connect the space frame and the body frame\footnote{Define the space frame simply equal to body frame at some specific time, for instance, at the initial time.}. So we have the following coordinate transformation on absolute configuration space $Q$ of the three-particle system
\[\left(\begin{array}{c}x_{1}\\ y_{1}\\ z_{1}\\ x_{2}\\ y_{2}\\ z_{2}\\ x_{3}\\ y_{3}\\ z_{3}\end{array}\right)\rightarrow \left(\begin{array}{c}x_{cm}\\ y_{cm} \\z_{cm} \\\alpha \\\beta \\ \gamma \\\lambda\\s_{1}\\ s_{2}    \end{array}\right)\]
From our previous discussions and after some calculations, we have derived the expression of the mass metric \textbf{M} on $Q$ in these new coordinates. The result is as follows\newline
\begin{equation}\label{3bmminnewcor}
dl^{2}= \frac{m_{3}(m_{1}+m_{2})\lambda^{2} sin^{2}(s_{2})}{m_{1}m_{2}sin^{2}(s_{1}+s_{2})}ds_{1}^{2}
\end{equation} 
\[+ \frac{m_{3}(m_{1}+m_{2})\lambda^{2} sin^{2}(s_{1})}{m_{1}m_{2}sin^{2}(s_{1}+s_{2})}ds_{2}^{2}\]

\[+\big(1+\frac{m_{2}}{m_{1}}+\frac{m_{3}(m_{1}+m_{2})sin^{2}(s_{2})}{m_{1}m_{2}sin^{2}(s_{1}+s_{2})}  \big)d\lambda^{2}\]\[+\sum_{a,b}A_{ab}\omega^{a}\omega^{b}\]\[+(m_{1}+m_{2}+m_{3})(dx_{cm}^{2}+dy_{cm}^{2}+dz_{cm}^{2})\]
Using the above line element, one can express the infinitesimal increment of Newton's absolute time $dt$ in terms of the system's motion (infinitesimal increments of particles spatial positions), i.e. (see also \cite{8},\cite{10}) 
\begin{equation}\label{et} 
dt=\frac{dl}{\sqrt{E-V}}
\end{equation} 
It is the increment of ephemeris time for this three-particle universe. Now it becomes clear that for a spatially larger three-particle system by a factor $b>1$, the same amount of relational motion $(ds_{1},ds_{2})$ will result \footnote{For simplicity, consider the case where all collective momenta(linear, angular, and dilatational) are vanishing.} in a longer increment of ephemeris time\footnote{Longer by the same factor by which the spatial size of the system has been multiplied.} $dt$, when distances are measured with the fixed (non-scalable) absolute rod (unit of length) attached to Newton's absolute space (which, by the way, can also be used to measure and communicate changes of $\lambda$). Considering the ticks of the clocks as a certain amount of relational motion of the system (where the clock itself is part of), the relation between the seconds of the clocks after and before the spatial scale transformation of the system $\pmb{x}_{i}\rightarrow b\pmb{x}_{i}$ becomes $T\rightarrow T'=bT$. Of course, this difference in the rate of the clock ticking can only be sensed if we have access to the absolute Newtonian clock, which is unaffected by anything that happens to matter in the universe. This is also in complete agreement with our earlier discussions (see Section (II) of \cite{Me4} and \cite{8}) on the relation between the behavior of Planck's time unit under the global spatial scale transformations $\pmb{x}_{i}\rightarrow b \pmb{x}_{i}$, namely $T_{p}\rightarrow T'_{p}=b T_{p}$. This relation was derived there directly from the principle of relationalism.\newline
  
As can be seen from (\ref{siminla}), the Lagrangian of the three-particle system in the center of mass frame is the following function on $Q_{cm}$ 
\begin{equation}\label{l3pa} 
\mathscr{L}=\frac{1}{2}\frac{m_{3}(m_{1}+m_{2})\lambda^{2} sin^{2}(s_{2})}{m_{1}m_{2}sin^{2}(s_{1}+s_{2})}\dot{s}_{1}^{2}
\end{equation}
\[+ \frac{1}{2}\frac{m_{3}(m_{1}+m_{2})\lambda^{2} sin^{2}(s_{1})}{m_{1}m_{2}sin^{2}(s_{1}+s_{2})}\dot{s}_{2}^{2}\]
\[+\big(1+\frac{m_{2}}{m_{1}}+\frac{m_{3}(m_{1}+m_{2})sin^{2}(s_{2})}{m_{1}m_{2}sin^{2}(s_{1}+s_{2})}\big)\dot{\pmb{\lambda}}^{2}\]
\[+\frac{1}{2}\sum_{a,b}A_{ab}\Omega^{a}\Omega^{b}-V\]
where as before $\dot{\pmb{\lambda}}=\frac{\dot{\lambda}}{\lambda}$, and
\begin{equation}\label{3bp} 
V=G\big(\frac{m_{1}m_{2}}{\mid\pmb{x}_{2}-\pmb{x}_{1}\mid}+\frac{m_{1}m_{3}}{\mid\pmb{x}_{3}-\pmb{x}_{1}\mid}+\frac{m_{2}m_{3}}{\mid\pmb{x}_{3}-\pmb{x}_{2}\mid}\big)
\end{equation}
Although the actual scale ($\lambda$) of the system appears explicitly in the above Lagrangian, it is, in fact, a scale invariant Lagrangian if one takes the transformation of the unit of time and the gravitational constant after the performance of a global scale transformation\footnote{As mentioned earlier, in a larger universe the clocks tick slower, and $G$ becomes larger (in MNT)} into account. Here we elaborate on this point more. As for the units used in (\ref{l3pa}), one has measured lengths with respect to Newton's absolute rod (with respect to which the diameter of our system happens to be $\lambda$), and one uses the ephemeris unit of time\footnote{which is a kind of internal clock of the system with respect to which Newton's second law remains valid.} for time measurements. Now one sees the quantity $\lambda \dot{s}=\lambda \frac{ds}{dt_{e}}$, where subscript $e$ stands for ephemeris, is an invariant quantity under the system's spatial scaling $\pmb{x}_{i}\rightarrow \pmb{x}_{i}'=b\pmb{x}_{i}$, as such a transformation leads to \[\lambda\rightarrow \lambda'=b\lambda\] and \[\delta t_{e}\rightarrow \delta t_{e}'=b\delta t_{e}\] 
So one gets \[\lambda \dot{s}=\lambda \frac{ds}{dt_{e}}\rightarrow \lambda' \dot{s}'=\lambda' \frac{ds}{dt_{e}'}=b\lambda \frac{ds}{b.dt_{e}}=\lambda \frac{ds}{dt_{e}}= \lambda \dot{s}\] 
To be realistic about length measurements and incorporate this realism into physical theory, one must use a unit of length that consists of matter. Thus, an internal (or relational) unit of length must be used instead of the invisible absolute Newtonian unit of length, which has its origin and justification in absolute space. For example, take the system's diameter as an internal unit of length. This means $\lambda=1$. For an increment of the system's shape $(ds_{1},ds_{2})$, we calculate the increment of time using the formula for the increment of ephemeris time (\ref{et}), in which we also set $\lambda=1$, i.e., $dt=dt_{e}\mid_{\lambda=1}$. Then the expression of Lagrangian (\ref{l3pa}) with the use of relational time and length becomes as follows
\begin{equation}\label{detleflikeit} 
\mathscr{L}=\frac{1}{2}\frac{m_{3}(m_{1}+m_{2}) sin^{2}(s_{2})}{m_{1}m_{2}sin^{2}(s_{1}+s_{2})}\dot{s}_{1}^{2}
\end{equation}
\[+ \frac{1}{2}\frac{m_{3}(m_{1}+m_{2}) sin^{2}(s_{1})}{m_{1}m_{2}sin^{2}(s_{1}+s_{2})}\dot{s}_{2}^{2}\]
\[+\big(1+\frac{m_{2}}{m_{1}}+\frac{m_{3}(m_{1}+m_{2})sin^{2}(s_{2})}{m_{1}m_{2}sin^{2}(s_{1}+s_{2})}\big)\dot{\pmb{\lambda}}^{2}\]
\[+\frac{1}{2}\sum_{a,b}A_{ab}\Omega^{a}\Omega^{b}-V(s_{1},s_{2})\]
where now its scale invariance becomes explicitly visible.\newline The third term here, which contains the scale velocity $\dot{\pmb{\lambda}}$, is to be interpreted with caution. It has the same origin as the fourth term in the Lagrangian. The fourth term causes the Coriolis and centrifugal forces in a rotating reference frame (as in a body frame of a rotating system). Similarly, the third term causes dilational forces in a spatially expanding reference frame (as in a body frame of an expanding system with an internal unit of length). Note also that the scale velocity $\dot{\pmb{\lambda}}=\frac{\dot{\lambda}}{\lambda}$ has the dimension of the inverse of time, as it is the ratio of the change of the system's scale$\delta \lambda$ during one second of internal time $\delta t_{e}$ to the system's scale$\lambda$.
At an instant of time $t_{0}$, when one is viewing the expanding system from some point of absolute space, one can manually set the absolute length unit equal to the instantaneous scale of the system at that time and express$\dot{\pmb{\lambda}}$
in this new absolute unit of length. It simply becomes $\dot{\pmb{\lambda}}=\dot{\lambda}=\frac{\delta \lambda}{\delta t_{e}\mid_{\lambda=1}}$, where during the small observation time interval $[t_{0},t_{0}+\delta t_{e}\mid_{\lambda=1}]$ we have fixed the absolute length unit to $\lambda\mid_{t=t_{0}}$, and therefore can measure the new scale of the system at the end of the time interval $\lambda\mid_{t=t_{0}+\delta t_{e}\mid_{\lambda=1}}=\lambda\mid_{t=t_{0}}+\delta\lambda$.\newline 

The equations of motion of the two shape degrees of freedom $s^{1}$, and $s^{2}$ for the case of vanishing total angular velocity $\omega_{t}=0$ \footnote{This is always the only relevant physical situation if the system under consideration is supposed to represent our universe. It follows from the fact that the frame built from the background stars and galaxies is an inertial reference frame.} can be given using (\ref{shad})
\[\frac{d}{dt}\big(\frac{\partial \mathscr{L}}{\partial \dot{s}^{i}}\big)-(\frac{\partial}{\partial s^{i}})^{*}\mathscr{L}=0\]
After some lengthy calculations, we end up with the following coupled second-order nonhomogeneous non-linear differential equations for the shape degrees of freedom of a nonrotating three-body system
\[
sin^{2}(s_{2})sin(s_{1}+s_{2})\ddot{s}_{1}-3sin^{2}(s_{2})cos(s_{1}+s_{2})\dot{s}_{1}^{2}+\]\[2sin(s_{2}\big(cos(s_{2})sin(s_{1}+s_{2})-sin(s_{2})cos(s_{1}+s_{2})\big)\dot{s}_{2}\dot{s}_{1}\]\[+2\dot{\pmb{\lambda}}sin(s_{1}+s_{2})sin^{2}(s_{2})\dot{s}_{1}\]\[+sin(s_{1})\big(cos(s_{1})sin(s_{1}+s_{2})-sin(s_{1})cos(s_{1}+s_{2})\big)\dot{s}_{2}^{2}\]
\begin{equation}\label{3EL1} 
+2\dot{\pmb{\lambda}}^{2}sin^{2}(s_{2})cos(s_{1}+s_{2})+\frac{m_{1}m_{2}}{m_{3}(m_{1}+m_{2})}\frac{\partial V}{\partial s_{1}}=0 
\end{equation} 
 and
\[
sin^{2}(s_{2})sin(s_{1}+s_{2})\ddot{s}_{2}\]\[+cos(s_{1}+s_{2})\big(sin^{2}(s_{1}-2sin^{2}(s_{2})\big)\dot{s}_{2}^{2}\]\[+2sin(s_{2})\Big(cos(s_{2})sin(s_{1}+s_{2})\]\[-sin(s_{2})cos(s_{1}+s_{2})\Big)\dot{s}_{1}\dot{s}_{2}\]\[+2\dot{\pmb{\lambda}}sin^{2}(s_{2})sin(s_{1}+s_{2})\dot{s}_{2}\]\[+sin(s_{2})\Big(cos(s_{2})sin(s_{1}+s_{2})\]\[-sin(s_{2})cos(s_{1}+s_{2})\Big)\dot{s}_{1}^{2}\]\[+2\dot{\pmb{\lambda}}^{2}sin(s_{2})\Big(cos(s_{2})sin(s_{1}+s_{2})\]\[-sin(s_{2})cos(s_{1}+s_{2})\Big)\]
\begin{equation}\label{3EL22} 
-\frac{m_{1}m_{2}}{m_{3}(m_{1}+m_{2})}\frac{\partial V}{\partial s_{2}}=0
\end{equation}
If, additionally, the system is non-expanding, i.e., $\dot{\pmb{\lambda}}=0$, the reduced equations of motion on shape space become as follows
\[sin^{2}(s_{2})sin(s_{1}+s_{2})\ddot{s}_{1}-3sin^{2}(s_{2})cos(s_{1}+s_{2})\dot{s}_{1}^{2}\]\[+2sin(s_{2})\bigg(cos(s_{2})sin(s_{1}+s_{2})\]\[-sin(s_{2})cos(s_{1}+s_{2})\bigg)\dot{s}_{2}\dot{s}_{1}\]\[+sin(s_{1})\bigg(cos(s_{1})sin(s_{1}+s_{2})\]\[-sin(s_{1})cos(s_{1}+s_{2})\bigg)\dot{s}_{2}^{2}\]
\begin{equation}\label{3EL1n} 
+\frac{m_{1}m_{2}}{m_{3}(m_{1}+m_{2})}\frac{\partial V}{\partial s_{1}}=0 
\end{equation}
and
\[sin^{2}(s_{2})sin(s_{1}+s_{2})\ddot{s}_{2}\]\[+cos(s_{1}+s_{2})\big(sin^{2}(s_{1}-2sin^{2}(s_{2})\big)\dot{s}_{2}^{2}\]\[+2sin(s_{2})\bigg(cos(s_{2})sin(s_{1}+s_{2})\]\[-sin(s_{2})cos(s_{1}+s_{2})\bigg)\dot{s}_{1}\dot{s}_{2}\]\[+sin(s_{2})\bigg(cos(s_{2})sin(s_{1}+s_{2})\]\[-sin(s_{2})cos(s_{1}+s_{2})\bigg)\dot{s}_{1}^{2}\]
\begin{equation}\label{3EL2n} 
-\frac{m_{1}m_{2}}{m_{3}(m_{1}+m_{2})}\frac{\partial V}{\partial s_{2}}=0
\end{equation}
\newline  
Now we need to discuss the structure of the potential (\ref{3bp}) in more detail. As mentioned elsewhere(\cite{8},\cite{Me4}), the potential function can be considered to be the product of two functions, $G$, and $f$, on the absolute configuration space $Q$, i.e., $V=Gf$. The form of the function $f$ is known from the time of Isaac Newton, and in the special case of our three-body system, it can be read from (\ref{3bp}). In contrast to $f$, the $E(3)$ invariant function $G$ has remained unknown to this day. All we know about $G$ is that it now has a value of about $6.67384(80)\times 10^{-11} m^{3}.kg^{-1}.s^{-2}$ on and near Earth, with a relative uncertainty of $2\times 10^{-5}$, making it by far the least precisely known natural constant. One candidate function for $G$ is
\begin{equation}\label{ffG} 
G:=\sqrt{I_{cm}}=\sqrt{\sum_{i=1}^{N}m_{i}\mid\pmb{x}_{i}-\pmb{x}_{cm}\mid^{2}}
\end{equation}
which satisfies the requirement of being a Euclidean invariant function and being homogeneous of first-degree under scale transformations.
For the region \[0\leq sin(s_{1}),sin(s_{2})<sin(s_{1}+s_{2})\] on the three-body shape space, one has \[\lambda=\mid \pmb{x}_{2}-\pmb{x}_{1}\mid\] 
A lift of the shape $s=(s_{1},s_{2})$ to the absolute configuration space can be given as follows
\[\pmb{x}_{1}=\left[\begin{array}{c}0\\ 0\\ 0\end{array}\right],\pmb{x}_{2}=\left[\begin{array}{c}\lambda\\ 0\\ 0\end{array}\right],\pmb{x}_{3}=\left[\begin{array}{c}\frac{sin(s_{2})cos(s_{1})}{sin(s_{1}+s_{2})}\lambda\\ \frac{sin(s_{2})sin(s_{1})}{sin(s_{1}+s_{2})}\lambda\\ 0\end{array}\right]\]
for which the location of the center of mass becomes as follows 
\[\pmb{x}_{cm}= \left[\begin{array}{c}\frac{m_{2}\lambda}{m_{1}+m_{2}+m_{3}}+\frac{m_{3}sin(s_{2})cos(s_{1})}{(m_{1}+m_{2}+m_{3})sin(s_{1}+s_{2})}\lambda\\ \frac{m_{3}sin(s_{2})sin(s_{1})}{(m_{1}+m_{2}+m_{3})sin(s_{1}+s_{2})}\lambda\\ 0\end{array}\right]\]
Now by putting all these back in (\ref{ffG}), and after some  calculations, one gets the following expression for $G$
\[G=\lambda\bigg(\frac{sin^{2}(s_{2})}{sin^{2}(s_{1}+s_{2})}\big(m_{3}+m_{3}^{2}(1-\frac{2}{M})\big)\]\[-2\frac{m_{2}m_{3}}{M}\frac{sin(s_{2})cos(s_{1})}{sin(s_{1}+s_{2})}+m_{2}(1-\frac{m_{2}}{M})\bigg)^{1/2}\]
and for $f$
\[f=\lambda^{-1}\bigg(m_{1}m_{2}+m_{1}m_{3}\frac{sin(s_{1}+s_{2})}{sin(s_{2})}\]\[+\frac{m_{2}m_{3}}{\frac{sin^{2}(s_{2})}{sin^{2}(s_{1}+s_{2})}-2\frac{sin(s_{2})cos(s_{1})}{sin(s_{1}+s_{2})}+1}\bigg) \]
Finally, the potential function $V=Gf$ of our three-body system in new coordinates takes the following form
\[V=\bigg(m_{1}m_{2}+m_{1}m_{3}\frac{sin(s_{1}+s_{2})}{sin(s_{2})}\]\[+\frac{m_{2}m_{3}}{\frac{sin^{2}(s_{2})}{sin^{2}(s_{1}+s_{2})}-2\frac{sin(s_{2})cos(s_{1})}{sin(s_{1}+s_{2})}+1}\bigg)\]\[\times\bigg(\frac{sin^{2}(s_{2})}{sin^{2}(s_{1}+s_{2})}\big(m_{3}+m_{3}^{2}(1-\frac{2}{M})\big)\]\[-2\frac{m_{2}m_{3}}{M}\frac{sin(s_{2})cos(s_{1})}{sin(s_{1}+s_{2})}+m_{2}(1-\frac{m_{2}}{M})\bigg)^{1/2}\]
which now manifests its scale invariance explicitly.\newline 
By using this function in the two reduced Euler-Lagrange equations (\ref{3EL1n}), (\ref{3EL2n}), we finally obtain the reduced equations of motion on shape space of the three-body system. \newline\newline
For the simplest case where the system under consideration is neither rotating nor expanding w.r.t. the absolute space, the Lagrangian function becomes 
\begin{equation}\label{relational3bpl}
\mathscr{L}=\frac{1}{2}\frac{m_{3}(m_{1}+m_{2}) sin^{2}(s_{2})}{m_{1}m_{2}sin^{2}(s_{1}+s_{2})}\dot{s}_{1}^{2}
\end{equation}
\[+ \frac{1}{2}\frac{m_{3}(m_{1}+m_{2}) sin^{2}(s_{1})}{m_{1}m_{2}sin^{2}(s_{1}+s_{2})}\dot{s}_{2}^{2}\] 
\[+V(s_{1}s_{2})\]

\section{Conclusion}
In this paper, the derivation of the Euler-Lagrange equations of motion in non-holonomic frames, known as the Boltzmann-Hamel equations, and the methods for their reduction to the internal configuration space $Q_{int}=\frac{Q}{E(3)}$ of the classical mechanics(\cite{1},\cite{2},\cite{3},\cite{7},\cite{9},\cite{99},\cite{6})\footnote{The reduced equations of motion are also known as Lagrange-Poincare(\cite{LPE},\cite{MarsdenRatiuScheurle},\cite{MestdagCrampin}) equations of motion.}, has been extended to include the entire similarity group $Sim(3)$, with the help of which we have derived the reduced equations of motion of the modified Newtonian theory (introduced in \cite{Me4} and\cite{8}) on shape space $S=\frac{Q}{Sim(3)}$.\newline
We first constructed representations of the group $Sim(3)$ and its Lie algebra $\textbf{sim}(3)$ on $Q$, and discussed how a vector on shape space can be lifted horizontally into the center of mass configuration space $Q_{cm}$, constructed the new connection form $\omega_{s}$ for the $Sc$ fiber bundle and showed that this connection form is flat. As a consequence of the latter, quotienting out the configuration space w.r.t. the group of scale transformations $Sc$ leads to no additional curvature in the resulting base space. Thus, the curvature in shape space is caused solely by the quotienting w.r.t. the group of rotations $SO(3)$. Furthermore, we explained how the action $\mathcal{A}$ of $Sc$ on the absolute phase space of the modified Newtonian theory, is used to derive the unique metric $\textbf{N}$ on the $Sim(3)$-reduced tangent bundle $\frac{T(Q)}{\mathcal{A}_{Sim(3)}}$. On the other hand, we have discussed that by considering the role of the rulers in determining the geometry of space, the measured mass metric is itself scale invariant, so again there is no arbitrariness involved in the metric of shape space. In this regard, we have explained among other things, the relationship between the choice of a length unit and the choice of a conformal factor(introduced in \cite{DSN} for the purpose of derivation of a metric on shape space), and elaborated that all reasonable choices of length units lead to the same metric on shape space.\newline\newline
Using these new ingredients, we have derived the reduced equations of motion (\ref{shad}) of an $N$-particle system for its shape degrees of freedom, whose behavior in absolute space and time is given by the \textit{modified Newtonian mechanics}. The principle of relationalism guarantees, among other things, the existence of the laws of motion on shape space in a simple way. As the simplest non-trivial example of the extended formalism, we have explicitly derived the equations of motion (\ref{3EL1}),(\ref{3EL22}) for the shape degrees of freedom of a three-particle system.\newline

\textbf{Acknowledgement:} We would like to express our sincere gratitude to Prof. Dr. Detlef Dürr for his exceptional supervision and valuable guidance throughout this research project. We greatly appreciate his open-minded approach to physics and his invaluable contributions. We also thank Prof. Dr. Roderich Tumulka and Prof. Dr. Peter Pickl for their greate supervision, generous support, and numerous enlightening discussions.

\section{Appendix}
\subsection{Mass tensor}
Originally the Kinetic energy $K$ of a classical $N$-particle system is expressed(defined) in Cartesian coordinates as following
\begin{equation}\label{KEN}
K=\frac{1}{2}\sum_{i=1}^{N}m_{i}\dot{\textbf{x}}_{i}^{2}=\frac{1}{2}[\dot{\textbf{x}}_{1},...,\dot{\textbf{x}}_{N}]\textbf{M} \left[\begin{array}{c}\dot{\textbf{r}}_{1}\\ ...\\ \dot{\textbf{x}}_{N}\end{array}\right]
\end{equation} where $\dot{\textbf{x}}_{i}:=\frac{d\textbf{x}_{i}}{dt}$ with $\textbf{x}_{i}=\left[\begin{array}{c}x_{3i-2}\\ x_{3i-1}\\ x_{3i}\end{array}\right]$ and $\textbf{M}$ is the so called mass matrix which is in this case just a block diagonal $3N\times 3N$ matrix with $\begin{bmatrix}m_{j} & 0 & 0 \\0 & m_{j} & 0\\ 0 & 0 & m_{j} \end{bmatrix}$ as its $j$'s block. 
\newline 
Here as usual the configuration space is coordinatized by $x_{1},x_{2},...,x_{3N}$ which are in turn the collection of Cartesian coordinates $x_{3i-2}, x_{3i-1}, x_{3i}$ used to denote the position(vector in $\mathbb{R}^{3}$) of the $i$'th particle $\textbf{r}_{i}$.\newline 
Now if the system suffers a number of holonomic constraints, the generalized coordinates $q_{1},q_{2},...,q_{f}$ (with $f<3N$ standing for total number of remaining degrees of freedom), can be used for coordinatizing the new (generalized) configuration space.\newline
Now if one rewrites the kinetic energy $K$ in terms of this new generalized coordinates $q_{j}$ and their velocities $\dot{q}_{j}$ one ends up usually with a much more complicated expression than \ref{KEN} where it was in fact coordinate independent, and a simple diagonal quadratic form in the velocities. In generalized coordinates, it is quadratic but not necessarily homogeneous in the velocities $\dot{q}_{j}$, and has an arbitrary dependence on the coordinates $q_{j}$ (through $\textbf{M}$).\newline
If the coordinate transformation between the set of Cartesian coordinates ${x_{1},...,x_{3N}}$ and the generalized coordinates ${q_{1},...,q_{f}}$ is time-independent (see \cite{12}), the kinetic energy is written as
\begin{equation}\label{kine}
K=\frac{1}{2}\sum_{k,l}M_{kl}\dot{q}_{k}\dot{q}_{l}=\frac{1}{2}[\dot{q}_{1},...,\dot{q}_{f}]\textbf{M}\left[\begin{array}{c}\dot{q}_{1}\\ ...\\ \dot{q}_{f}\end{array}\right]
\end{equation}  
Where $M_{kl}=\sum_{j=1}^{N}m_{j}\frac{d\textbf{x}_{j}}{dq_{k}}.\frac{d\textbf{x}_{j}}{dq_{l}}=\sum_{j=1}^{N}\sum_{i=0}^{2}m_{j}\frac{dx_{3j-i}}{dq_{k}}\frac{dx_{3j-i}}{dq_{l}} $ are elements of the $f\times f$ matrix $\textbf{M}$.  \newline
The Lagrangian of classical mechanics is shown to be $L=K-V$, where the potential $V$ is usually independent of the generalized velocities $\dot{q}_{i}$. The conjugate momentum to $q_{i}$ is defined as
\begin{equation}\label{momentum}
p_{i}=\frac{\partial L}{\partial q_{i}}=\frac{\partial K}{\partial \dot{q}_{i}}=\sum_{j=1}^{f}M_{ij}\dot{q}_{j}
\end{equation} 
thus the expression \ref{kine} for the kinetic energy which involved just the velocities can be rewritten as 
\begin{equation}\label{sub}
K=\frac{1}{2}\sum_{i=1}^{f}p_{i}\dot{q}_{i}
\end{equation} 

\subsection{Adjoint and Coadjoint action of a Lie-Group}
Let $G$ be a Lie group, and $\bf G$ it's Lie algebra, and $\bf G ^{*}$ be the dual vector space of $\bf G$. The adjoint representation of $g\in G$ on $\bf G$ is defined by
\begin{equation}\label{adjact}
   Ad_{g}(Y)=\frac{d}{dt}\mid_{t=0}(g e^{tY}g^{-1})
\end{equation}
for $Y \in \bf G$.\newline
The coadjoint action of $g\in G$ on $\bf G^{*}$ is characterized by 
\begin{equation}
    <Ad^{*}_{g}(\xi),Y>=<\xi , Ad_{g^{-1}}(Y)>
\end{equation}
for $\xi \in \bf G^{*}$\newline
Here, $<,>:\bf g^{*}\times \bf g \rightarrow R$ is the dual pairing.\newline
In summary: $ad_{g}(x)=gxg^{-1}$, $Ad_{g}=(ad_{g})_{*}:\bf G\rightarrow \bf G$ being called the adjoint action , $Ad^{*}_{g}:\bf G^{*}\rightarrow \bf G^{*}$ being called the coadjoint action. 
\newline
\subsection{Isomorphism $R$}
There exists an isoporphism $R$ between Lie-algebra $\textbf{so}(3)$ of rotation group, and the linear space $\wedge^{2}\mathbb{R}^{3}$ of all antisymmetric tensors of order $2$, which we want to explain shortly. \newline
Take ${\pmb{e}_{1},...\pmb{e}_{3}}$ as an orthonormal basis of $\mathbb{R}^{3}$. Then $\pmb{e}_{i}\wedge \pmb{e}_{j}$ with $i<j$ constitutes an orthonormal basis of $\wedge^{2}\mathbb{R}^{3}$.
The inner product in $\wedge^{2}\mathbb{R}^{2}$ is defined as the following
\begin{equation}\label{tvsp}
(u\wedge v \mid x\wedge y)= \begin{vmatrix}
(u\mid x) & (u\mid y) \\ 
(v\mid x) & (v\mid y)
\end{vmatrix}
\end{equation}
 
One can easily check that for two two-vectors (or tensors of order 2) $\xi=\sum_{i<j}\xi_{ij}\pmb{e}_{i}\wedge \pmb{e}_{j}$ and $\zeta=\sum_{k<l}\zeta_{kl}\pmb{e}_{k}\wedge \pmb{e}_{l}$, definition \ref{tvsp} leads to the following
\begin{equation}
(\xi\mid\zeta)=\sum_{i<j}\xi_{ij}\zeta_{ij}
\end{equation}

Now we identify the Lie-algebra of the rotation group in 3 dimensions \textbf{so}(3) with the space of two forms (anti-symmeric tensors) $\wedge^{2}\mathbb{R}^{3}$ by the isomorphism $R$ 
\begin{equation}\label{R}
R: \wedge^{2}\mathbb{R}^{3} \xrightarrow{\sim} \textbf{so}(3)
\end{equation}
\[\xi \rightarrow R_{\xi}\]
So for $u$,$v$,$x\in \mathbb{R}^{3}$ we define the following
\begin{equation}\label{isoR}
R_{u\wedge v}(x):=(v\mid x)u-(u\mid x)v
\end{equation}  
$R_{u\wedge v}$ is in fact a 3 dimensional square matrix and it's multiplication by a 3 dimensional vector $x$ is given by the last equation. For $\xi \in\wedge^{2}\mathbb{R}^{3}$ and $x=\sum x_{j}\pmb{e}_{j}\in \mathbb{R}^{3}$ One can also write the above formula as 
\begin{equation}
R_{\xi}(x)=\sum_{i}(\sum_{j}\xi_{ij}x_{j})\pmb{e}_{i}
\end{equation}
That is, $R_{\xi}$ is an antisymmetric matrix with entries $\xi_{ij}$.\newline
Given the natural scalar product of the Lie algebra; $(\alpha\mid\beta)=\frac{1}{2}tr(\alpha\beta^{T})$ for $\alpha$,$\beta \in \textbf{so}$(3) one can show that the identification $R$ is even an isometry from $\wedge^{2}\mathbb{R}^{d}$ to \textbf{so}(d).

As explained in \cite{5}, space $\wedge^{2}\mathbb{R}^{3}$ can be identified with $\mathbb{R}^{3}$ by $\pmb{e}_{1}\wedge \pmb{e}_{2}\rightarrow \pmb{e}_{3}$ and the cyclic permutations. Hence if one sets \[\xi_{12}=\phi^{3}, \xi_{23}=\phi^{1}, \xi_{31}=\phi^{2}\] the two vector \[\xi=\sum_{i<j}\xi_{ij}\pmb{e}_{i}\wedge \pmb{e}_{j}\] is identified with \[\phi=\sum \phi^{i} \pmb{e}_{i}\] 
So in this case in effect $R$ becomes a linear isomorphism from $\mathbb{R}^{3}$ to $\textbf{so}(3)$ i.e.
\[R: \mathbb{R}^{3}\rightarrow \textbf{so}(3)\]
\begin{equation}\label{isoR1} 
    R_{\xi}(x)=R_{\phi}(x)=-\phi\times x
\end{equation}
for $x\in\mathbb{R}^{3}$.\newline 
Alternatively, $R_{e_{1}}$ is the matrix $(\xi_{ij})$ with the only nonzero elements \newline$\xi_{23}=-\xi_{32}=1$.\newline   
One can also show(\cite{5}) that $R$ is Ad-equivariant i.e. $R_{g\phi}=Ad_{g}R(\phi)=gR(\phi)g^{-1}$.\newline 
There exists the following properties for the map $R$, and the inertial tensor $A_{x}$ 
\begin{subequations}\label{properties of R A}
\begin{align}
    R_{a}b=a\times b     \\
    R_{ga}=gR_{a}g^{-1}    \\
    R_{a}.R_{b}=<a\mid b> \\
    A_{gx}(a)=gA_{x}(g^{-1}a):= Ad_{g}A_{x}(a) \\
    (x\mid R_{\xi}y)=(x\wedge y\mid \xi)\\
    (R_{\xi}x\mid R_{\eta}y)=(R_{\xi}x\wedge y\mid \eta)
\end{align}
\end{subequations}
where $a,b\in \mathbb{R}^{3}$ and $g\in SO(3)$. 
\newpage
\subsection{Symbols}

      $t^{(\pmb{n})} $ :   Newton's absolute time \newline
      $x$ :    A point on $Q_{cm}:=\frac{Q}{\mathbb{R}^{3}}$ \newline
      $q$ :    A point on $Q_{int}=\frac{Q_{cm}}{SO(3)}$ \newline
       $s$ :    A point on shape space $S:=\frac{Q}{Sim(3)}$ \newline
       $\pmb{r}_{i}$ :  $i$'s Jacobi vector of an $N$-particle system\\ 
         $\lambda$ :   Scale variable of a system\\
         $\dot{\lambda}$ :   Scale velocity of a system\\
         $\dot{\pmb{\lambda}}$ :   Scale velocity of a system measured in internal units\\
       $\alpha,\beta,\gamma$ :   Euler angles connecting a body frame and the space frame\\
         $\{\textbf{e}_{1},\textbf{e}_{2},\textbf{e}_{3}\}$ :    Fixed laboratory frame, or space frame\\
       $\{\textbf{e}_{1}',\textbf{e}_{2}',\textbf{e}_{3}'\}$ :    Body frame\\
       $g$ :  Rotation which brings the space frame to the body frame\\
          $\textbf{J}=\sum_{\alpha=1}^{N}m_{\alpha}x_{\alpha}\times\frac{\partial}{\partial x_{\alpha}}$ :   Total angular momentum\\
          $\Omega^{a}$: Components of angular velocity in space frame\\
          $\Omega'^{a}$: Components of angular velocity in body frame\\
         $\textbf{J}$ :    $=\sum_{a=1}^{3}\textbf{e}_{a}J_{a}=\sum_{a=1}^{3}\textbf{e}_{a}'L_{a}$\\
         $J_{a}$ :   $=(\textbf{e}_{a}\mid \textbf{J})$\\ 
         $L_{a} :  = (\textbf{e}_{a}'\mid \textbf{J})$ Left invariant vector fields on $SO(3)$\\
          $J_{a}\textbf{r}_{i}$ :   $=\textbf{e}_{a}\times \textbf{r}_{i}$\\
           $L_{a}\textbf{r}_{i}$ :  $=\textbf{e}_{a}'\times\textbf{r}_{i}=g\big(\textbf{e}_{a}\times\sigma_{i}(q)\big)$\\
          $\omega^{a}(J_{b})$ :   $=\delta^{a}_{b}$ \\
          $\omega'^{a}(L_{b})=\theta^{a}(L_{b})$:  $=\delta^{a}_{b}$ \\
          $\theta^{a}:$ Left invariant one forms on $SO(3)$\\
         $g^{-1}dg :=\sum_{a=1}^{3}\theta^{a}R(\textbf{e}_{a})$\\
         $\psi^{a}:$ Right invariant 1-forms on $SO(3)$\\
         $dgg^{-1}=:\sum_{a=1}^{3}\psi^{a}R(\textbf{e}_{a})$\\
          $\pmb{k}$ :   Curvature tensor of shape space\\
          $\pmb{c}$ :    Speed of light\\
          $b$ :   Letter used to characterize scale transformations by a factor $b\in\mathbb{R}^{+}$\\
          $Sc$ :    Group of spatial scale transformations(of matter)\\
          $G_{rs}$ :    Group of spatial rotations and scale transformations\\
          $A$ : Moment of inertia tensor of a $N$-particle system\\
          $\textbf{A}$ : Gauge fields on $Q_{int}$\\
          $\mathcal{A}_{g}$ : Action of $g\in Sim(3)$ on the tangent bundle $T(Q)$ \\
          $\textbf{D}$ :    Dilational momentum operator\\
          $D$ :    Value of system's dilational momentum measured in internal units\\
          $\textbf{M}$ :    Mass metric on $Q_{cm}$ or $Q$\\
          $B$ :    Metric on $Q_{int}$\\
         $\textbf{N}$: Metric on the reduced tangent bundle$\frac{T(Q)}{\mathcal{A}_{Sim(3)}}$\\
          $N$ :    Metric on the shape space\\
         $\textbf{M}_{\lambda\lambda}$ :   Scale component of the mass metric in shape and scale coordinates. $a(s)$\\
         $\textbf{M}^{(m)}$ :    The measured mass metric on $Q$\\



\medskip

\bibliographystyle{unsrt}
\bibliography{main}
\end{document}